\documentclass[aps,prb,twocolumn,showpacs,longbibliography,superscriptaddress]{revtex4-1}
\usepackage[utf8]{inputenc}
\usepackage[T1]{fontenc}

\usepackage{graphicx}
\usepackage{color}
\usepackage{amsfonts}
\usepackage{amsmath}
\usepackage{bm}
\usepackage{amssymb}
\usepackage{dcolumn}
\usepackage{cases}
\usepackage{booktabs}
\usepackage{setspace}
\usepackage{array}
\usepackage{blindtext}
\usepackage{placeins}
\usepackage{physics}
\usepackage{upgreek}
\usepackage{epstopdf}

\newcommand{\FeCoGe}{Fe$_{1-y}$Co$_{y}$Ge~}
\newcommand{\FeCoGens}{Fe$_{1-y}$Co$_{y}$Ge\xspace}

\newcommand{\approxsymbol}{$\sim$}

\newcommand{\Tc}{$T_{\text{c}}$\xspace}
\newcommand{\Ms}{$M_{\text{s}}$\xspace}

\newcommand{\Xdc}{$\chi_{\text{dc}}$\xspace}
\newcommand{\Xac}{$\chi_{\text{ac}}$\xspace}

\newcommand{\pxy}{$\rho_{xy}$~}

\newcommand{\degC}{$^{\circ}$C\xspace}
\newcommand{\degrees}{$^{\circ}$~}

\newcommand{\mum}{$\upmu$m~}
\newcommand{\microamps}{$\upmu$A~}

\begin{document}
	
\title{Helical magnetic structure and the anomalous and topological Hall effects\\ in epitaxial B20 \FeCoGe films}
	
\author{Charles~S.~Spencer}
\affiliation{School of Physics and Astronomy, University of Leeds, Leeds LS2 9JT, United Kingdom}

\author{Jacob~Gayles}
\affiliation{Max Planck Institute for Chemical Physics of Solids, 01187 Dresden, Germany}
\affiliation{Institut f\"ur Physik, Johannes Gutenberg Universit\"at Mainz, D-55099 Mainz, Germany}

\author{Nicholas~A.~Porter}
\affiliation{School of Physics and Astronomy, University of Leeds, Leeds LS2 9JT, United Kingdom}

\author{Satoshi~Sugimoto}
\altaffiliation{Present address: Institute for Solid State Physics, University of Tokyo, 5-1-5 Kashiwa-no-ha, Kashiwa, Chiba 277-8581, Japan}
\affiliation{School of Physics and Astronomy, University of Leeds, Leeds LS2 9JT, United Kingdom}

\author{Zabeada~Aslam}
\affiliation{Leeds Electron Microscopy and Spectroscopy Centre, School of Chemical and Process Engineering, University of Leeds, Leeds LS2 9JT, United Kingdom}

\author{Christian~J.~Kinane}
\affiliation{ISIS Neutron and Muon Source, STFC Rutherford Appleton Laboratory, Chilton, Didcot, Oxon OX11 0QX, United Kingdom}

\author{Timothy~R.~Charlton}
\altaffiliation{Present address: Neutron Scattering Division, PO Box 2008 MS 6473, Oak Ridge National Lab, Oak Ridge, TN 37831-6473, U.S.}
\affiliation{ISIS Neutron and Muon Source, STFC Rutherford Appleton Laboratory, Chilton, Didcot, Oxon OX11 0QX, United Kingdom}

\author{Frank~Freimuth}
\affiliation{Peter Gr{\"u}nberg Institut \& Institute for Advanced Simulation, Forschungszentrum J{\"u}lich and JARA, 52425 J{\"u}lich, Germany}

\author{Stanislav~Chadov}
\affiliation{Max Planck Institute for Chemical Physics of Solids, 01187 Dresden, Germany}

\author{Sean~Langridge}
\affiliation{ISIS Neutron and Muon Source, STFC Rutherford Appleton Laboratory, Chilton, Didcot, Oxon OX11 0QX, United Kingdom}

\author{Jairo~Sinova}
\affiliation{Institut f\"ur Physik, Johannes Gutenberg Universit\"at Mainz, D-55099 Mainz, Germany}
\affiliation{Institute of Physics ASCR, v.v.i., Cukrovarnicka 10, 162 53 Praha 6 Czech Republic}

\author{Claudia~Felser}
\affiliation{Max Planck Institute for Chemical Physics of Solids, 01187 Dresden, Germany}

\author{Stefan~Bl{\"u}gel}
\affiliation{Peter Gr{\"u}nberg Institut \& Institute for Advanced Simulation, Forschungszentrum J{\"u}lich and JARA, 52425 J{\"u}lich, Germany}

\author{Yuriy~Mokrousov}
\affiliation{Peter Gr{\"u}nberg Institut \& Institute for Advanced Simulation, Forschungszentrum J{\"u}lich and JARA, 52425 J{\"u}lich, Germany}
\affiliation{Institut f\"ur Physik, Johannes Gutenberg Universit\"at Mainz, D-55099 Mainz, Germany}

\author{Christopher~H.~Marrows}\affiliation{School of Physics and Astronomy, University of Leeds, Leeds LS2 9JT, United Kingdom}
	
\date{\today}
	
\begin{abstract}
Epitaxial films of the B20-structure compound \FeCoGe were grown by molecular beam epitaxy on Si (111) substrates. The magnetization varied smoothly from the bulk-like values of one Bohr magneton per Fe atom for FeGe to zero for non-magnetic CoGe. The chiral lattice structure leads to a Dzyaloshinskii-Moriya interaction (DMI), and the films' helical magnetic ground state was confirmed using polarized neutron reflectometry measurements. The pitch of the spin helix, measured by this method, varies with Co content $y$ and diverges at $y \sim 0.45$. This indicates a zero-crossing of the DMI, which we reproduced in calculations using first principles methods. We also measured the longitudinal and Hall resistivity of our films as a function of magnetic field, temperature, and Co content $y$. The Hall resistivity is expected to contain contributions from the ordinary, anomalous, and topological Hall effects. Both the anomalous and topological Hall resistivities show peaks around $y \sim 0.5$. Our first principles calculations show a peak in the topological Hall constant at this value of $y$, related to the strong spin-polarization predicted for intermediate values of $y$. Our calculations predict half-metallicity for $y = 0.6$, consistent with the experimentally observed linear magnetoresistance at this composition, and potentially related to the other unusual transport properties for intermediate value of $y$. Whilst it is possible to reconcile theory with experiment for the various Hall effects for FeGe, the large topological Hall resistivities for $y \sim 0.5$ are much larger then expected when the very small emergent fields associated with the divergence in the DMI are taken into account.
\end{abstract}
	
\maketitle

\section{Introduction}

Topologically protected spin structures, termed skyrmions, have recently come to the forefront of spintronics research\cite{Nagaosa2013}. Skyrmions boast the possibility for high density magnetic non-volatile storage \cite{Fert2013} and the ability to be used in logic gates \cite{Zhang2015c,Zhang2015b}. These exotic magnetic systems host many of the novel spintronic phenomena such as the anomalous Hall effect \cite{Nagaosa2010}, the spin Hall effect \cite{Sinova2015}, and current-induced torques\cite{Schulz2012,Jonietz2010,Hals2014,Everschor2011}. They can be manipulated with smaller current densities than that of domain walls \cite{Jonietz2010}, with the possibility to be used in racetrack memory \cite{Parkin2015}.

\begin{figure}[t]
\includegraphics[width=8cm]{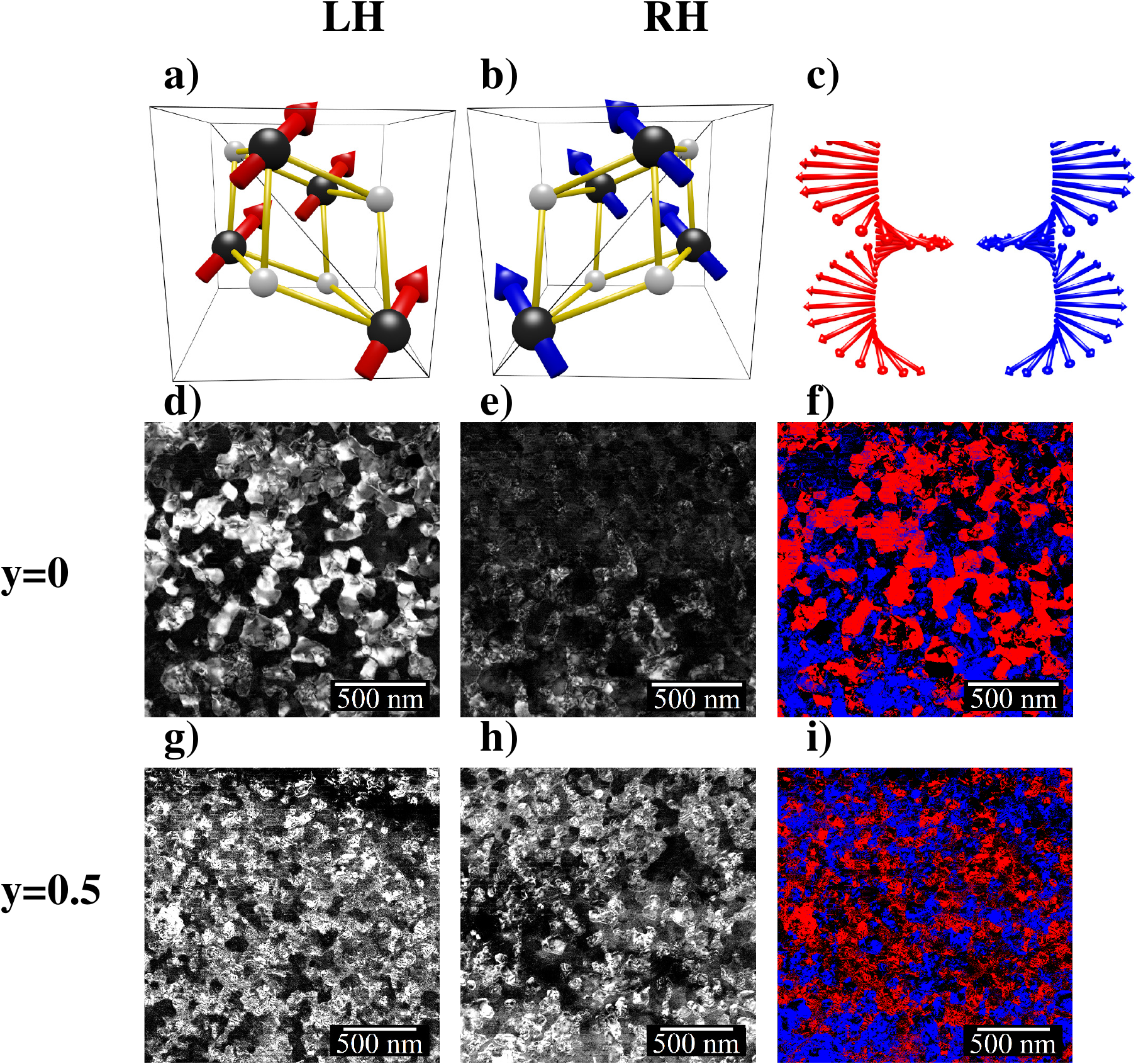}
\caption{(color online) Chiral crystal structure of \FeCoGens. B20 unit cell diagrams for (a) left-handed and (b) right-handed chiral crystals where the helical ground state propagates along the dashed line. The dark colored ion magnetic moments belong to Fe$_{1-y}$Co$_{y}$ and the light colored ions to Ge. (c) Diagram of magnetic helix structure for left-handed and right-handed chirality. (d-k) Transmission electron micrographs taken from \FeCoGe samples with $y =$ 0 and 0.5. Dark-field images from (d,g) ($1\bar{1}\bar{1}$) and (e,h) ($\bar{1}11$) reflections corresponding to the left-handed and right-handed chiral grains respectively. (f,i) False color composite of both dark-field images from each $y$ value showing the coverage of grains across the film. Red (blue) grains have left- (right-) handed crystal chirality.}
\label{chir}
\end{figure}

In the last decade there has been an emergence of research on many B20 compounds due to their possibility to stabilize skyrmionic crystals \cite{Muhlbauer2009a} where the topological Hall effect\cite{Bruno2004,Binz2008} (THE) can be observed\cite{Neubauer2009,Lee2009}. The inversion symmetry breaking in the crystal (see Fig.~\ref{chir}) and the presence of spin-orbit coupling (SOC) results in the so-called Dzyaloshinkii-Moriya interaction\cite{Dzyaloshinskii1957,Moriya1960} (DMI) manifesting itself, which prefers neighboring spins to point perpendicular to one another with a fixed chirality. Many B20 compounds display a helical ground state due to the competition of the DMI with the Heisenberg exchange\cite{Beille1983,Uchida2006}. This chiral magnetic texture drives ground state spin-currents\cite{Freimuth2016} that affects the observable transport properties in both the THE and the anomalous Hall effect (AHE). These emergent phenomena are theoretically connected through the Berry phase physics in real, momentum, and phase space, that correspond to the THE, AHE, and DMI respectively \cite{Freimuth2014}.

The B20 compounds have been extensively studied experimentally, however the connection between the Berry phases is still obscure. MnSi was the first B20 compound shown to display a stable skyrmion crystal \cite{Neubauer2009} below the magnetic transition temperature of 29~K. Experimental measurements of MnGe show the largest THE for any skyrmionic lattice \cite{Kanazawa2011}, which is due to the small skyrmion size of 3-6~nm. Furthermore, the THE does not depend on the chirality of Bloch skyrmions but changes sign and magnitude as a function of external field and as a function of temperature, suggesting a change in the skyrmion lattice\cite{Tanigaki2015} or shape \cite{Kanazawa2012}. The chirality of the skyrmion in the simplest case is determined by the sign of the DMI, where substituting Fe into MnGe caused a change in sign of the skyrmion at a critical concentration of $x = 0.8$ and the skyrmions texture becomes a trivial ferromagnet\cite{Shibata2013,Gayles2015,Kikuchi2016,Koretsune2015}. There have also been studies on thin films of FeGe, which show skyrmions close to room temperature with a helix pitch of 70~nm \cite{Yu2011} that can be driven at low current density \cite{Yu2012}. Topological Hall effects have been observed in a variety of different FeGe films \cite{Huang2012,Porter2014,Gallagher2017}, which can be discretized in small geometries comparable to the skyrmion size \cite{Kanazawa2015}. FeGe shows an AHE that is due primarily to the intrinsic mechanism \cite{Porter2014}. An enhanced ordering temperature has been seen in FeGe films grown on MgO \cite{Zhang2017}.

In addition to Mn$_{1-x}$Fe$_{x}$Ge, substituting Co into FeGe shows that the helix pitch also changes magnitude as a function of $y$ in polycrystalline samples of Fe$_{1-y}$Co$_{y}$Ge \cite{Grigoriev2014}. However, CoGe with the same crystal symmetry displays \textit{paramagnetism} in the ground state, with no net magnetization to observe AHE, THE, or any phenomena induced by DMI. We grew \FeCoGe films by molecular beam epitaxy (MBE) to observe the helical pitch and Hall effects as a function of Co concentration. As the magnetization is quenched as $y \rightarrow 1$ we also see the suppression of all spin-orbit effects. Our experimental results are directly compared with density functional theory (DFT) methods in two approximations of disorder, the virtual crystal approximation (VCA) and the coherent potential approximation (CPA) \cite{Soven1967,Taylor1967,Ebert2011,Mankovsky2017}.

We find experimentally that the substitution of Co into FeGe at concentration $y$ leads to changes in all the magnetic and magnetotransport properties. The magnetisation declines smoothly from roughly one Bohr magneton per Fe for $y=0$ to zero for non magnetic CoGe ($y=1$). The helix pitch diverges at $y \approx 0.45$, indicating a change of sign in the DMI, which is reproduced in our DFT results. Experimentally, we also find a range of of unusual transport properties for intermediate values of $y$, including peaks in the anomalous and topological Hall resistivities around $y \sim 0.5$. These are related to the very high degree of spin-polarisation we find theoretically for such values of $y$, including the prediction of a half-metallic state for $y=0.6$, with which the observation of a linear high-field magnetoresistance at that composition is consistent. Nevertheless, whilst the topological Hall constant is calculate to show a peak in this regime, the measured topological Hall resistivity is found to be much in excess of the theoretical upper limit based on the assumption of a fully dense skyrmion lattice.

In this report, we begin in section II with the experimental results, discussing the growth method, magnetometry, measurement of the helical magnetic structure and the transport measurements of the resistivity, the AHE and the THE. In the second part, section III, of this paper we discuss the electronic structure methods of full potential linearized augmented planewave (FLAPW) and spin polarized relativistic Korringa-Kohn-Rostoker (SPKKR), showing the calculations of the DMI, then the transport calculation of the AHE and the THE. We make a comparison of the experiment and the computational results for our substituted systems in section IV, before briefly concluding in section V.
	
\section{Experiment}	

\subsection{Sample growth and structural characterization}

Molecular beam epitaxy (MBE) was used to grow B20 \FeCoGe films on Si (111) substrates with room-temperature resistivity of 3-5 k$\Omega$ cm. The substrates were annealed at 1200\degC for 2 minutes to remove the native oxide and achieve a $(7 \times 7)$ surface reconstruction to ensure a clean and well ordered surface. This was verified \textit{in situ} by reflection high-energy electron diffraction (RHEED) and low-energy electron diffraction (LEED) and the substrate was allowed to cool before deposition to $< 50$\degC. The \FeCoGe films were grown by co-deposition from individual Fe, Co, and Ge electron beam sources and quartz crystal monitors were used to measure and regulate the flux from each source. To start the growth, approximately 1~nm of material (thickness of film before crystallization) was deposited. The sample was then heated to 230\degC to allow the layer to crystallize, forming a seed layer for growth. A further four 1~nm layers were then deposited with intervals of 15 minutes in-between each layer to allow for crystallization, which was verified by RHEED. The films were then co-evaporated with a net rate between 0.3-0.6 \AA /s using RHEED to monitor the structure during growth. After deposition a second LEED image was taken to verify $(111)$-oriented epitaxial growth. Finally after cooling to room temperature a cap layer of Ge was deposited to prevent oxidation.

\begin{figure}[t]
\includegraphics[width=8cm]{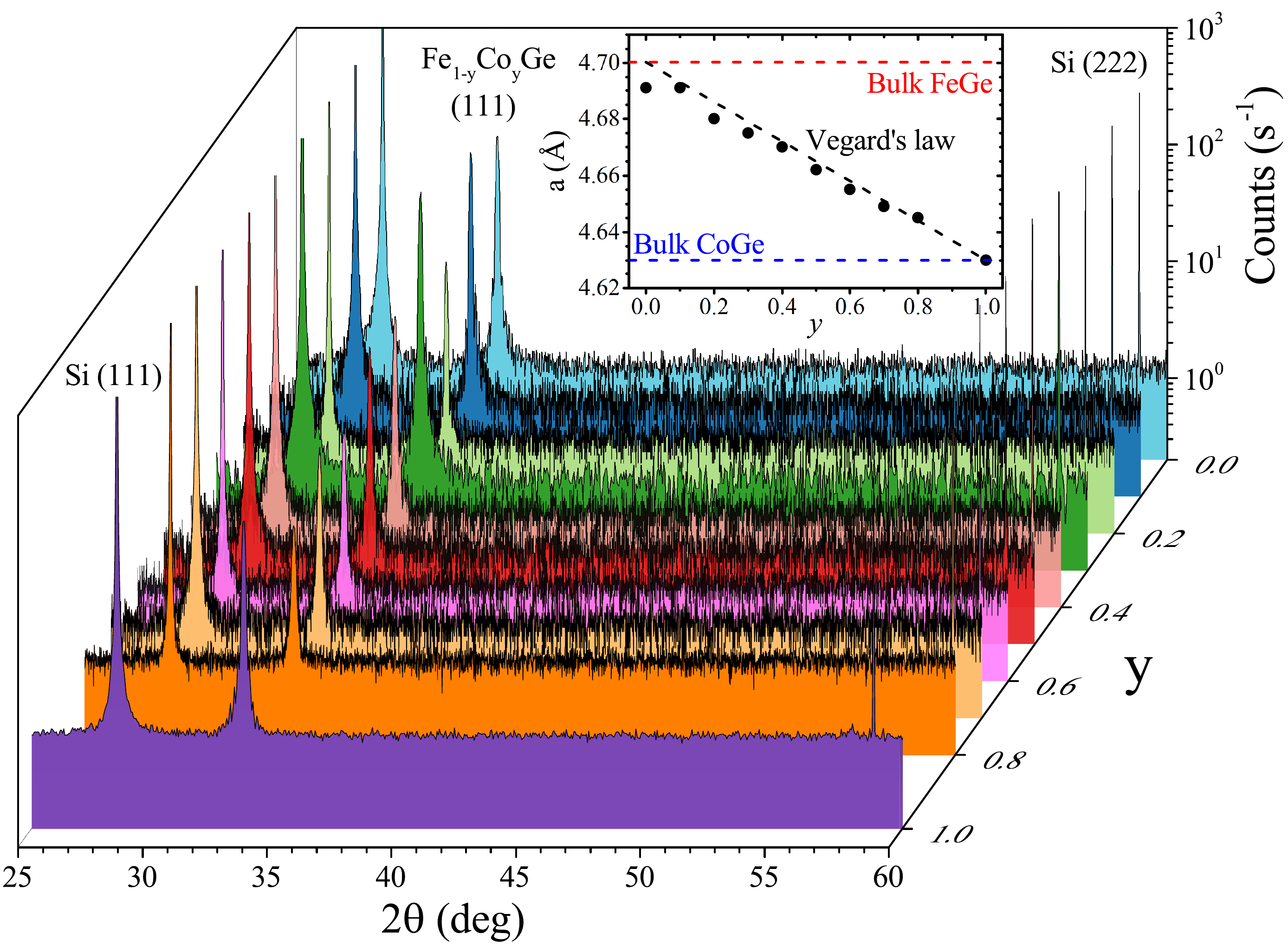}
\caption{(color online) X-ray diffraction data for the \FeCoGe films for all concentrations of $y$. Spectra have been offset for clarity. Inset: lattice constant $a$ measured from the position of the B20 \FeCoGe $(111)$ peak. \label{expXRD}}
\end{figure}

To verify the crystal phases present in the films, x-ray diffraction (XRD) measurements with Cu $K_\alpha$ radiation were used. The XRD spectra for each film composition $y$ is shown in Fig.~\ref{expXRD}. A single peak in the vicinity of $2\theta = 33$\degrees indicates the B20 $(111)$ reflection in \FeCoGens. The additional peaks correspond to the Si substrate $(111)$ and $(222)$ reflections respectively. A single peak for the film demonstrates the epitaxial and single phase character of our films.

The inset in Fig.~\ref{expXRD} shows the lattice constant $a$ as a function of $y$ determined from the position of the \FeCoGe $(111)$ peak assuming that cubic crystal symmetry is retained. For FeGe, $a = 0.4691 \pm 0.0001$~nm which is approximately 0.2\% less than bulk value of 0.4700~nm \cite{Lebech1989} and compares well with other reported films \cite{Huang2012,Porter2014}. This is due to the excellent lattice match between FeGe and Si. In Fig.~\ref{expXRD} the bulk value of 0.4631~nm for CoGe \cite{Takizawa1988} is matched to within the uncertainy by the measured value of $a = 0.4630 \pm 0.0001$~nm. Between these end members there is a good agreement with Vegard's law as $y$ varies, showing a consistent B20 crystal structure throughout the group. We conclude that our films have close to cubic lattice vectors, due to the small strain ($< -0.5\%$), which decreases with concentration $y$.

\begin{figure}[t]
\includegraphics[width=8cm]{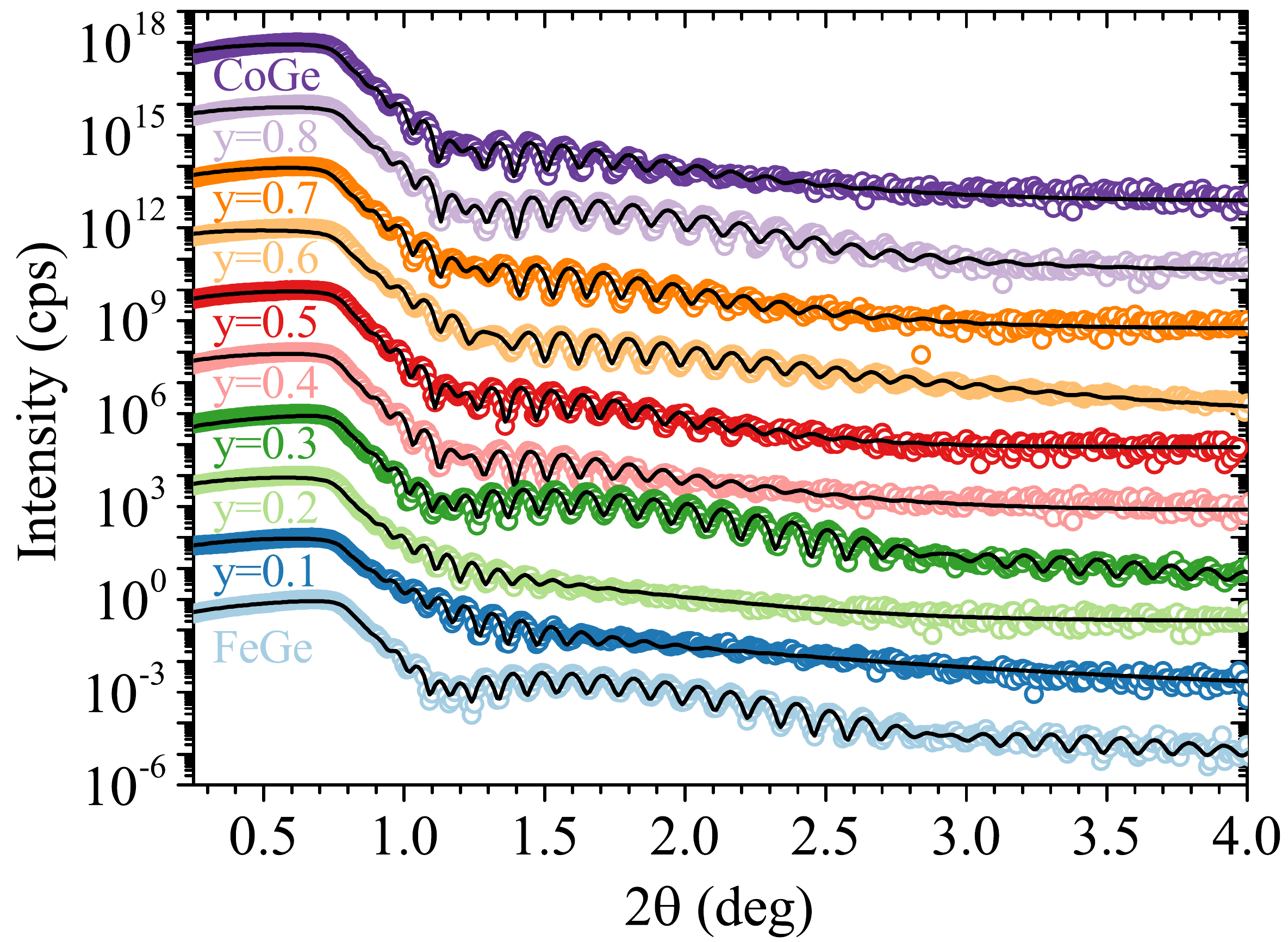}
\caption{(color online) X-ray reflectometry data for \FeCoGe (open symbols) and fits (line). Data sets have been offset for clarity.}
\label{expXRR}
\end{figure}

The film thickness and layer structure were examined using x-ray reflectometry (XRR), also with Cu $K_\alpha$ radiation. Each film was grown to a nominal thickness of 70~nm with a 4~nm Ge cap. In Fig.~\ref{expXRR} the measured XRR data (open circles) and fits (solid line) are shown. Each fit, performed using the GenX software \cite{Bjorck2007}, showed the films model a single continuous \FeCoGe layer and cap  which shows uniformity throughout the film. For each film the thickness $t$ was found to be within 10\% of the nominal value, with a top surface roughness of between 1 and 2~nm. A summary of the measured values for $a$ and $t$ for each sample is given in Table~\ref{tab_s1} in Appendix \ref{App_A}.

Plan-view transmission electron microscopy (TEM) was used to examine the chiral grain structure found in the films produced. The images were taken using a FEI Titan Themis 300 operated at 300~kV and collected using a Gatan OneView 16 Megapixel CMOS digital camera. For films grown on Si (111) neither grain chirality is favoured and the film is expected to be composed of both left-handed (LH) and right-handed (RH) chiral grains \cite{Karhu2010,Karhu2011,Porter2013,Porter2015}. The grains belonging to each chirality can be identified by using dark-field TEM, as shown in Ref. \onlinecite{Karhu2010}, here we align to the $[321]$ zone-axis of the film and the resulting dark-field images from the ($\bar{1}11$) (RH) and ($1\bar{1}\bar{1}$) (LH) reflections are shown for $y = 0$ and 0.5 in Fig.~\ref{chir} (d,e) and (g,h) respectively. In each image the grains with associated chirality appear bright and the two images from each chirality form an interlocking pattern showing the grains cover the film with apparently even size (\approxsymbol 100-200~nm) and probability. False color images produced using the LH and RH images for $y = 0$ and 0.5 are shown in Fig.~\ref{chir}(f) and (i) respectively.

\subsection{Magnetometry}
\begin{figure}[t]
\includegraphics[width=8cm]{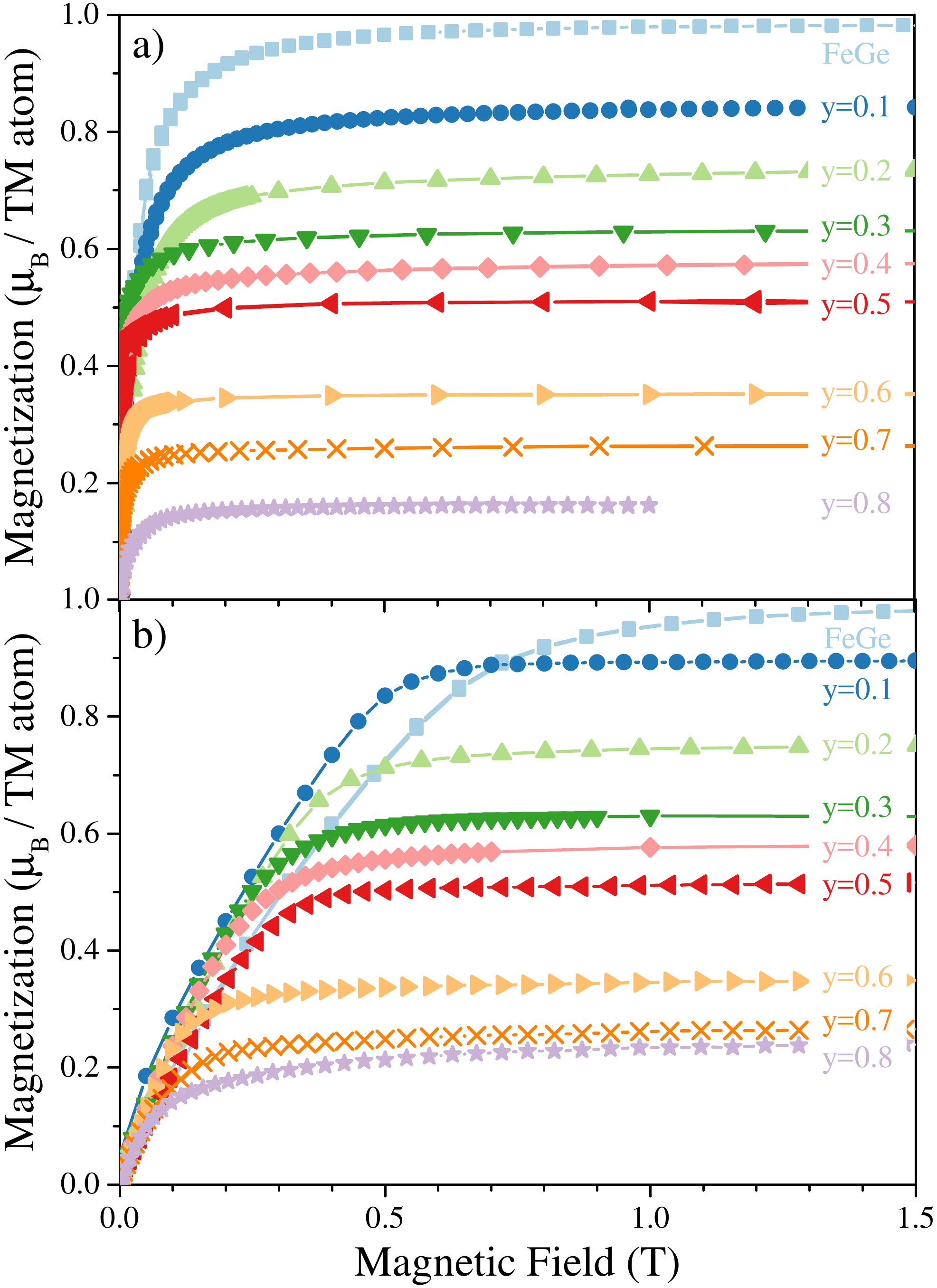}
\caption{(color online) Magnetization for (a) in-plane and (b) out-of-plane applied fields at 5~K (data shown for $y =$ 0.5 in-plane is at 10 K). \label{expMH}}
\end{figure}

The magnetic properties of the films were characterized using a superconducting quantum interference device vibrating sample magnetometer (SQUID-VSM). The temperatures were varied from 5~K up to room temperature with applied fields up to $\pm 6$~T. The magnetization for each sample with fields applied in-plane (IP) \FeCoGe [110] and out-of-plane (OOP) \FeCoGe [111] are shown in Fig.~\ref{expMH}(a) and (b) respectively. All films showed easy-plane anisotropy.

\begin{figure}[t]
\includegraphics[width=8cm]{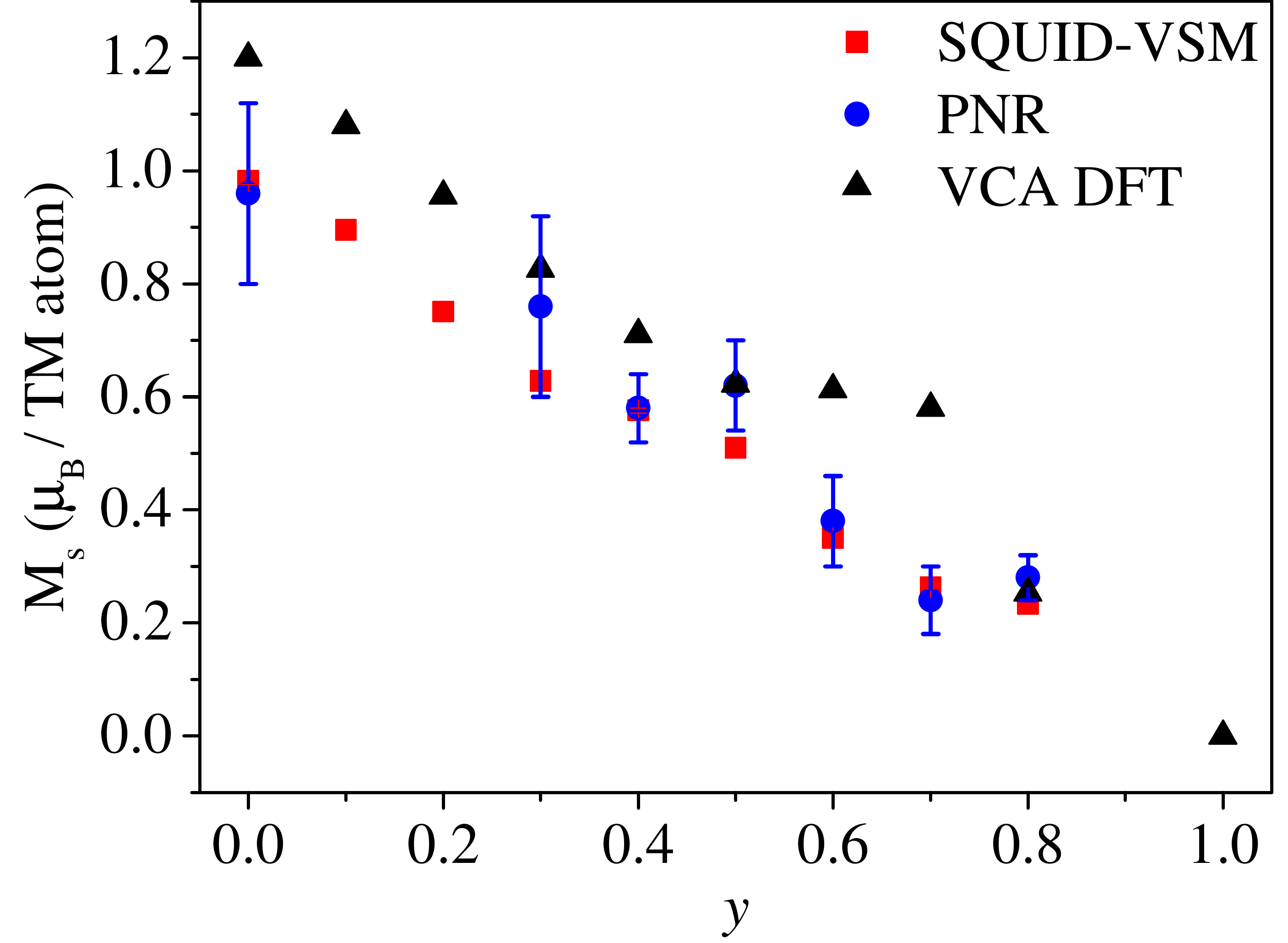}
\caption{(color online) Experimental values of saturation magnetization \Ms as a function of $y$ for measurements taken using SQUID-VSM (squares) and PNR (circles), and the results of VCA DFT calculations (triangles).}
\label{expMs}
\end{figure}

The in-plane saturation magnetization \Ms at 5~K, determined from these hysteresis loops at high field, is given in Fig.~\ref{expMs} as a function of $y$. For FeGe a magnetization of $360 \pm 10$~kA/m was found, which corresponds to a moment per Fe atom of $0.982 \pm 0.007$~$\mu_B$. This agrees with previously measured bulk \cite{Lebech1989} and thin film values \cite{Huang2012,Porter2012,Porter2015}. For the \FeCoGe samples the \Ms is found to decrease with increasing $y$ which is consistent with previous bulk measurements\cite{Grigoriev2015}. The density functional theory (DFT) results compared to experimental values show an over estimation of the moments due to exchange-correlation approximation and the half-metallic nature at $y = 0.6$ and 0.7, which will be discussed in subsequent sections.

\begin{figure}[t]
\includegraphics[width=8cm]{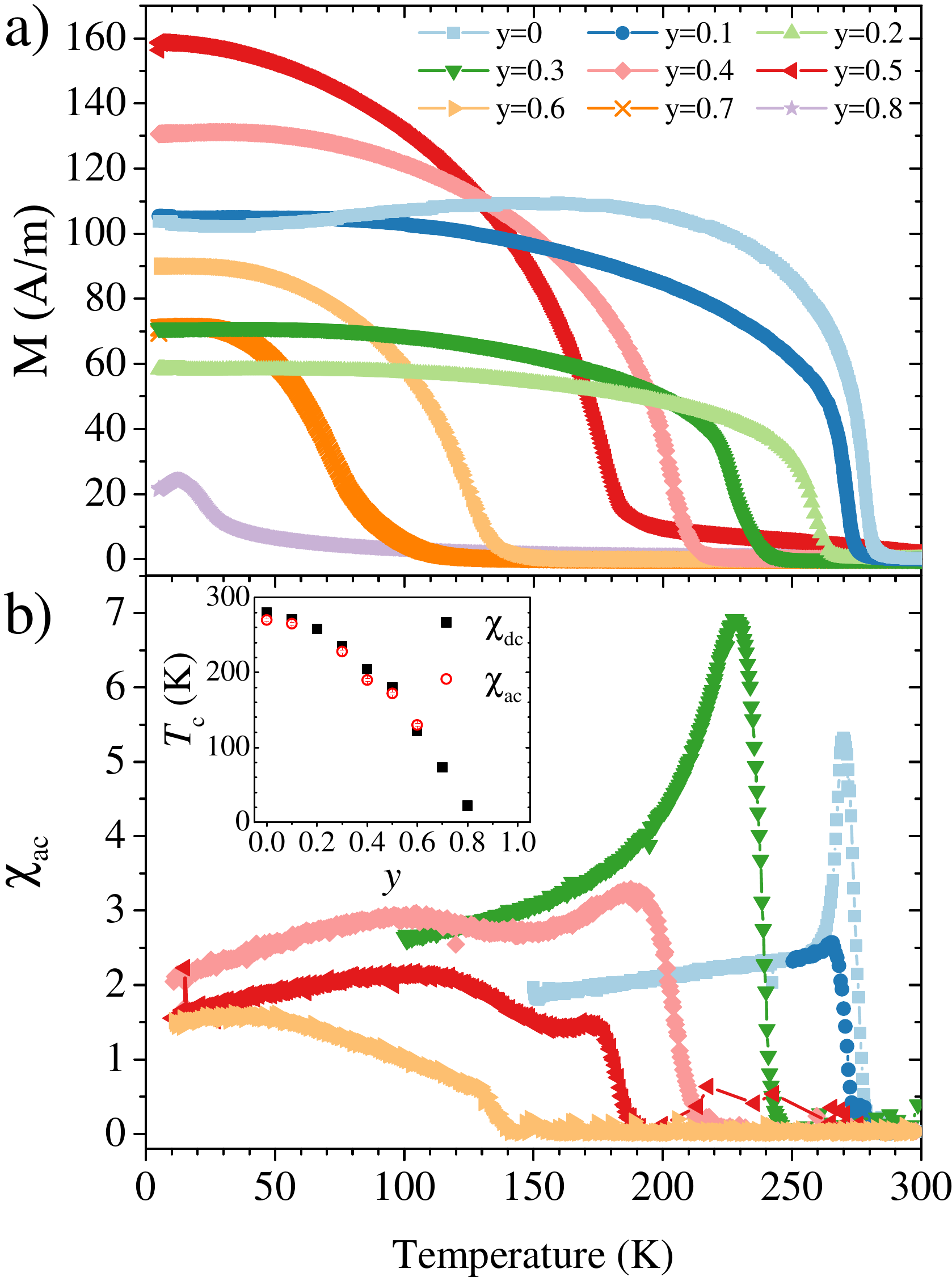}
\caption{(color online) Magnetization and magnetic susceptibility measurements for \FeCoGe as a function of temperature. (a) Magnetization $M$ for each $y$ at 10 mT. (b) \Xac for $y = 0$ to 0.6 with a static field at 2~mT and AC field at 1~mT at 23~Hz. Inset: Magnetic ordering temperature \Tc taken using \Xdc (squares) and \Xac (circles) as a function of $y$. \label{expTc}}
\end{figure}

To determine the magnetic ordering temperature \Tc we used both DC and AC susceptibility (\Xdc, \Xac) measurements. The results for each sample are shown in Fig.~\ref{expTc} with (a) and (b) showing the magnetization and \Xac measurements as a function of temperature respectively. For \Xdc an IP field of 10~mT was applied at room temperature and the moment was measured as the temperature was swept down to 5~K. The \Tc value was estimated from the peak found in $dM/dT$ at the onset of magnetic ordering. For \Xac the same temperature procedure was used with an applied IP static field of 2~mT and an AC field of 1~mT at 23~Hz. From these measurements \Tc was determined from the initial peak as temperature was decreased. For FeGe a sharp peak can been seen at the onset of magnetic ordering and as $y$ increases the peak becomes broader and a more complex behaviour develops. For FeGe the \Tc is found to be 280 $\pm 2$~K, which is again bulk-like \cite{Lebech1989}. The inset in Fig.~\ref{expTc}(b) shows the ordering temperatures and reveals a monotonic dependence on the concentration $y$, which agrees with previous experiments \cite{Grigoriev2015}.

\subsection{PNR Studies of the Helical Magnetic Structure}

To investigate the helical magnetic structure of these films we used polarized neutron reflectometry (PNR), which allows the magnetic depth profile of a film to be determined \cite{Blundell1992}. The PNR measurements of our compounds were taken at ISIS using PolRef, a time-of-flight reflectometer. Nominally 20 mm $\times$ 20 mm samples were mounted in a helium flow cryostat. A magnetic field was applied normal to the scattering plane and parallel to the \FeCoGe [110] direction (in the plane of the sample). A beam of polarized neutrons were then reflected from the sample and the neutron spin-up ($I_+$) and spin-down ($I_-$) reflected intensities were measured as a function of out-of-plane scattering vector $q_z = (4\pi/\lambda)\sin\theta$, where $\theta$ is the incident angle and $\lambda$ is the wavelength of the incident neutron. The spread of neutron velocities and two values of $\theta$ were used to provide a range for $q_z$ of  0.01 - 0.15 \AA$^{-1}$. As there is limited information at the higher wavevector transfers and for experimental expediency the PNR data presented here is shown with a range of $q_z$ up to 0.1~\AA$^{-1}$. Finally, the data were rebinned to a constant resolution of $\Delta q_z/q_z$ of 3\% consistent with the selected measurement resolution.
	
Before studying the helical structure, it was important to verify the chemical structure and to check the magnetic properties of the films. First, each sample was measured at room temperature (above \Tc) and at the maximum available field (667~mT) to determine the structure of the sample without any magnetic component and to compare this with the XRR results. The PNR values for \FeCoGe layer thickness were found to agree within 3\% of the XRR values (see Table~\ref{tab_s1} in Appendix~\ref{App_A}). The sample was then cooled to below \Tc and another measurement was taken to obtain the saturation moment. The \Ms values found using PNR are shown in Fig.~\ref{expMs} and agree well with the SQUID-VSM data to within error.

To observe the helical magnetic structure in these films, the method shown by Monchesky et al. in MnSi \cite{Karhu2012,Wilson2013,Wilson2014} was used. The magnetic depth profile provided by PNR is the sample averaged component of magnetization in the field direction as a function of depth $z$ analogous to a transverse spin density wave. Due to the growth method used, there is no influence on the chirality of the film produced, and so this results in the inevitable presence of chiral twinning in B20 thin films \cite{Karhu2011,Karhu2012}, which contain left- and right-handed helix structures (see Fig.~\ref{chir}). This leads to cancellation of the moments perpendicular to the applied field, and thus the measured depth profile is a 2D representation of the helix structure. As an applied field is increased this profile becomes distorted into a helicoid \cite{Wilson2014,Porter2015}. The magnetization profile, using this helicoid model, is given by:
\begin{eqnarray}\label{eqn:helicoid}
M(z) & = & M_0 + M_{\text{1}}\sin\left(\frac{2\pi z}{\Lambda_{\text{h}}} + \phi\right) + \nonumber \\ & & M_{\text{2}}\cos^2\left(\frac{2\pi z}{\Lambda_{\text{h}}} + \phi\right),
\end{eqnarray}
where $M$ is the magnetization, $M_0$ is an offset of the magnetization, $M_1$ and $M_2$ are fitting parameters, $\Lambda_{\text{h}}$ is the helical wavelength, and $\phi$ is a fitting parameter allowing adjustment of the phase of the helicoid. This profile was used in conjunction with the GenX software\cite{Bjorck2007} to fit the data. To fit the magnetic structure, the structural fitting parameters that were determined at room temperature were kept constant and only the magnetic parameters in Eq.~\ref{eqn:helicoid} were altered.
	
\begin{figure}[t]
\includegraphics[width=8cm]{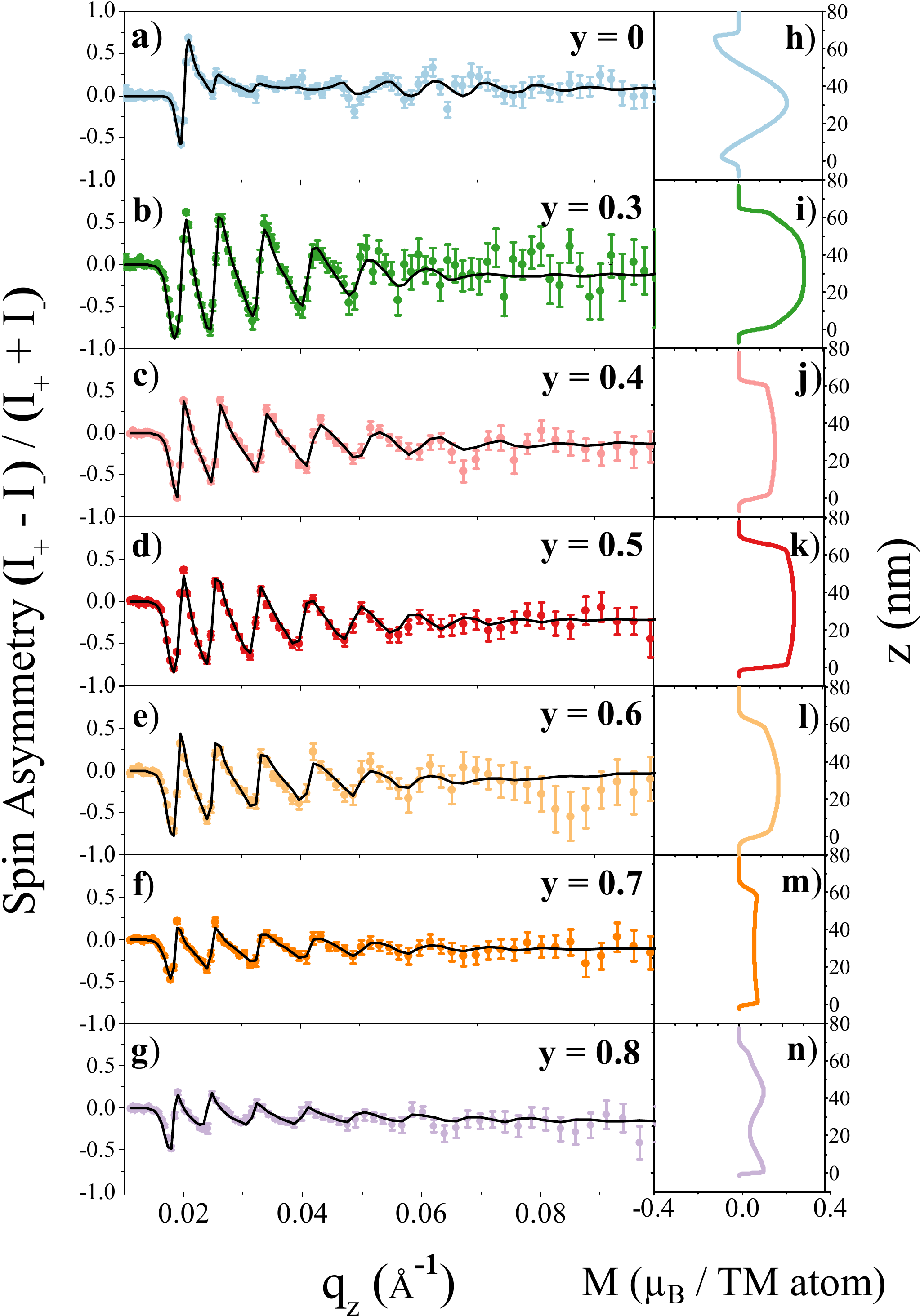}
\caption{(color online) PNR results for \FeCoGe. (a-g) Spin asymmetry data (circles) and fit (line) for films with varying $y$. (h-n) Magnetic depth profiles of each film leading to the corresponding fit.}
\label{expPNR}
\end{figure}

The samples were field cooled in a field of 5~mT to 50~K for $y = 0$ - 0.6 and 5~K for $y = 0.7$ and 0.8. Once cooled, the field was reduced to 1~mT (the smallest possible field that maintains the polarization of the neutron beam) to reduce distortion of the magnetic profile, before the measurements were taken. The results are shown in Fig.~\ref{expPNR}, plotted as a spin asymmetry (SA) defined as $(I_+ - I_-)/(I_+ + I_-)$, although the simultaneous fits were to the separate $I_+$ and $I_-$ curves. The data are shown as points and fits as solid lines, shown in the left panels (a-g), and the magnetic depth profiles that led to the fits are shown on the right in (h-n).
		
From these fits the helix wavelength $\Lambda_{\text{h}}$ was measured and the extracted values for each film are shown later in Fig.~\ref{DMIresults}, where they will be discussed in detail in comparison with the results of \textit{ab-initio} calculations. Nevertheless, a brief qualitative discussion is useful at this point. For FeGe the bulk value of $\Lambda_\mathrm{h} \approx 70$~nm, motivating our choice for a nominal film thickness $t$ of the same value. The FeGe film can indeed be seen to contain one whole turn of the helix. Thus, the helix pitch that the film has is the same value as in the bulk. Measurements of $k_\mathrm{h} = 2 \pi / \Lambda_\mathrm{h}$ in a series of \FeCoGe bulk crystals by small angle neutron scattering revealed a sharp minimum with $k_\mathrm{h} \sim 0$ for $y \sim 0.6$ \cite{Grigoriev2015}, implying another zero-crossing for the DMI at that composition. Our PNR data are consistent with this picture: the magnetic depth profile is almost uniform around that value of $y$.

\subsection{Magnetotransport}

To study the electron transport properties, Hall bar devices were fabricated using ultraviolet light photolithography. An 8-contact bar of 20 \mum width and 10 \mum contact spacing was chosen. The measurements presented here were taken using a DC current reversal method at $\pm$100 \microamps for temperatures from 5~K up to 300~K, which is above the magnetic ordering temperature for all values of $y$. This setup allows for measurement of the temperature dependence of both the longitudinal resistivity $\rho_{xx}$ and transverse (Hall) resistivity $\rho_{xy}$ simultaneously, along with any variation with field.

Ferromagnetic materials with non-trivial spin textures give rise to three contributions to the Hall resistivity \cite{Nagaosa2010}:
\begin{equation}\label{halleffect}
	\rho_{xy}= \rho_{xy}^\mathrm{OHE}+\rho_{xy}^\mathrm{AHE}+\rho_{xy}^\mathrm{THE}.
\end{equation}
The first contribution, from the ordinary Hall effect (OHE), $\rho_{xy}^\mathrm{OHE} = R_0 \mu_0H$, arises due to the Lorentz force and is directly dependent on the external magnetic field $H$. The second term, $\rho_{xy}^\mathrm{AHE} = R_\mathrm{s} \mu_0 M_z$, is due to the anomalous Hall effect (AHE) and is proportional to the component of magnetization along the field direction $M_z$ through the anomalous Hall coefficient $R_\mathrm{S}$ (Ref.~\onlinecite{Nagaosa2010}). The third term, $\rho_{yx}^\mathrm{THE} = R_{xy}^\mathrm{THE}B_\mathrm{eff}$, is the topological Hall effect (THE), which arises from Berry phases due to topologically non-trivial spin textures appearing as an effective magnetic field $B_\mathrm{eff}$, transversely accelerating electron quasiparticles with opposite spins in opposite directions\cite{Bruno2004,Tatara2008}. $B_\mathrm{eff}$ is determined by the skyrmion winding number density, $B_\mathrm{eff}=\Phi_0\frac{1}{4\pi}\int\hat{\mathbf m}\cdot(\partial_x\hat{\mathbf m}\times\partial_y\hat{\mathbf m})$, and $\Phi_0=\frac{h}{e}$. This has previously been detected in FeGe, MnGe, and MnSi, both in films and bulk \cite{Neubauer2009,Kanazawa2015,Kanazawa2011,Li2013,Franz2014}. In principle there are other contributions in the OHE that can lead to a nonlinearity in the Hall signal\cite{Porter2014}. In this work we do not focus on the OHE, but (along with the resistivity) on the AHE and THE, which arise from momentum space and real space Berry curvature, respectively.

\subsubsection{Resistivity and Magnetoresistance}

\begin{figure}[t]
\includegraphics[width=8cm]{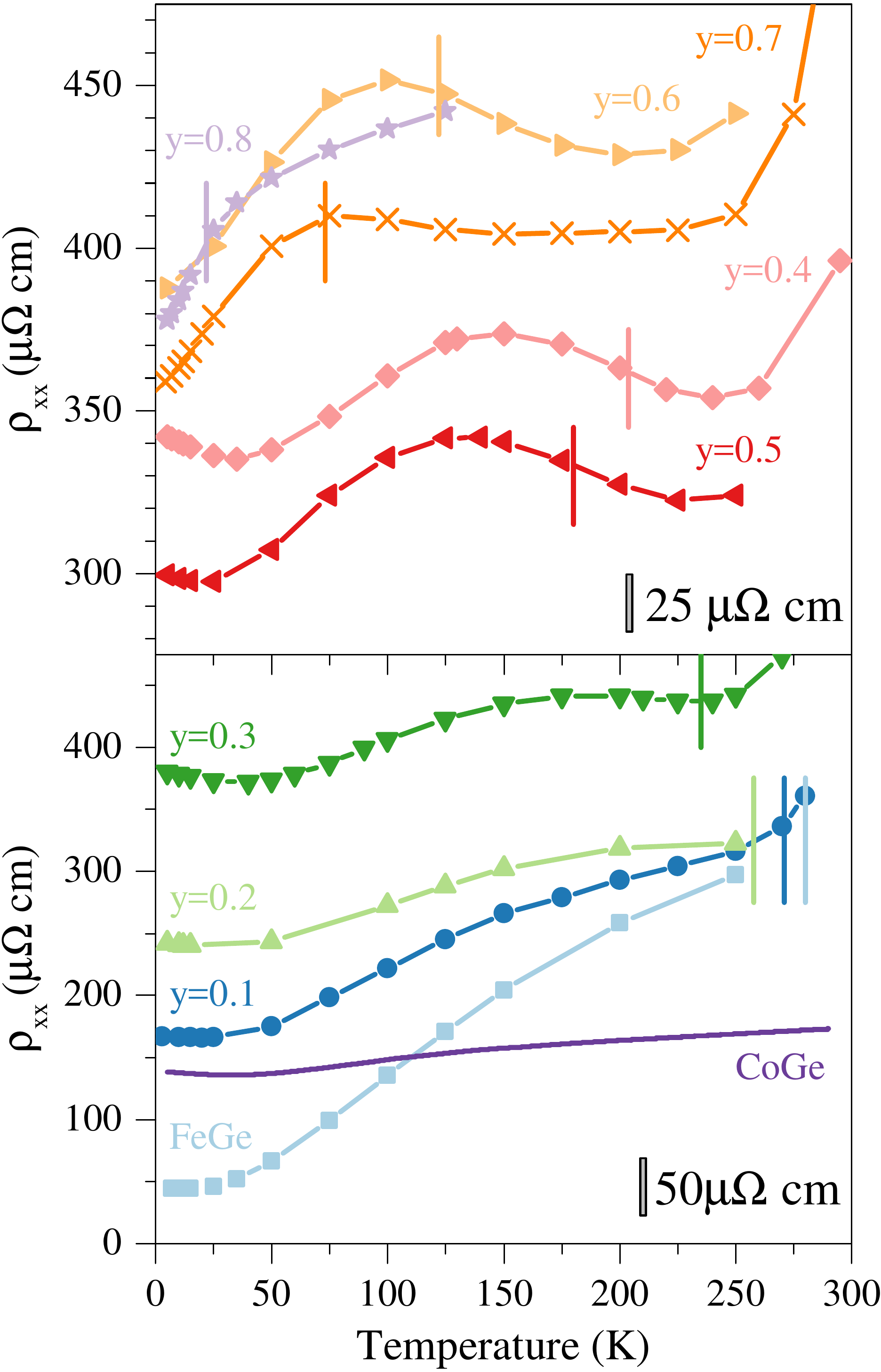}
\caption{(color online) Temperature dependent resistivity $\rho_{xx} (T)$ at zero magnetic field for all concentrations $y$. The data are separated into two panels to highlight details in $\rho_{xx}(T)$. The bottom panel shows $y = 0$ to 0.3 and $y = 1$ and top panel shows $y = 0.4$ to 0.8. Vertical lines show \Tc for respective concentration $y$ as determined from the data in Fig. \ref{expTc}. Data points for $0 \leq y \leq 0.8$ show measurements taken at fixed temperature and lines are guide for the eye, data shown for $y = 1$ was taken using a sweeping temperature. Note change in scale between panels. \label{exppxx}}
\end{figure}

The temperature dependent longitudinal resistivity $\rho_{xx} (T)$ at zero field is shown in Fig.~\ref{exppxx} for each composition $y$. The vertical line on each graph shows the ordering temperature $T_\mathrm{c}$ for that composition. FeGe ($y = 0$) and CoGe ($y = 1$) both show metallic behavior with resistivity rising with temperature. However the intermediate compositions all deviate and show a broad peak superimposed on this rise. For most intermediate values of $y$ the peak is large enough to make $\rho_{xx} (T)$ non-monotonic. The temperature at which the peak occurs is related to the magnetic ordering temperature, which appears at or close to an inflection point where $d^2 \rho_{xx} / d T^2 = 0$. Solid vertical lines are used to show the value of $T_\mathrm{c}$ for each value of $y$ determined from the susceptibility data shown in Fig.~\ref{expTc}. $T_\mathrm{c}$ coincides with an inflection point on the falling edge of the peak $\rho_{xx} (T)$. The deviation from the metallic behavior in the Co-substituted compounds in the intermediate range of Co concentration can be attributed to spin disorder, which is apparent in the DFT band structure calculations shown in Appendix C.

Anomalies in the resistivity are expected upon approaching $T_\mathrm{c}$ and have been shown for ferromagnetic materials to be caused by additional scattering due to spin fluctuations near the critical temperature \cite{rossiterbook}. Fisher and Langer \cite{Fisher1968} showed that in the case of short-range interactions the derivative of the resistivity, $d\rho_{xx}/d T$, should vary as the magnetic specific heat, which has been seen to be the case in MnSi \cite{Stishov2007,Bauer2013}. Whilst FeGe shows a conventional cusp in the specific heat\cite{Wilhelm2016}, there is no corresponding peak in $d\rho_{xx}/d T$ here or in previous work on thin films \cite{Huang2012,Porter2014}. An alternative theoretical approach is to look at the long-range interaction contribution, which has been used to describe the resistivity anomalies found in periodic magnetic structures, with spin-spiral rare-earth systems given as an example \cite{Suezaki1969}. In this description, a peak in the resistivity at a temperature below $T_\mathrm{c}$ is predicted under certain conditions, which is what we observe in our data with the most pronounced effect at $y = 0.5$.

\begin{figure}[t]
\includegraphics[width=8cm]{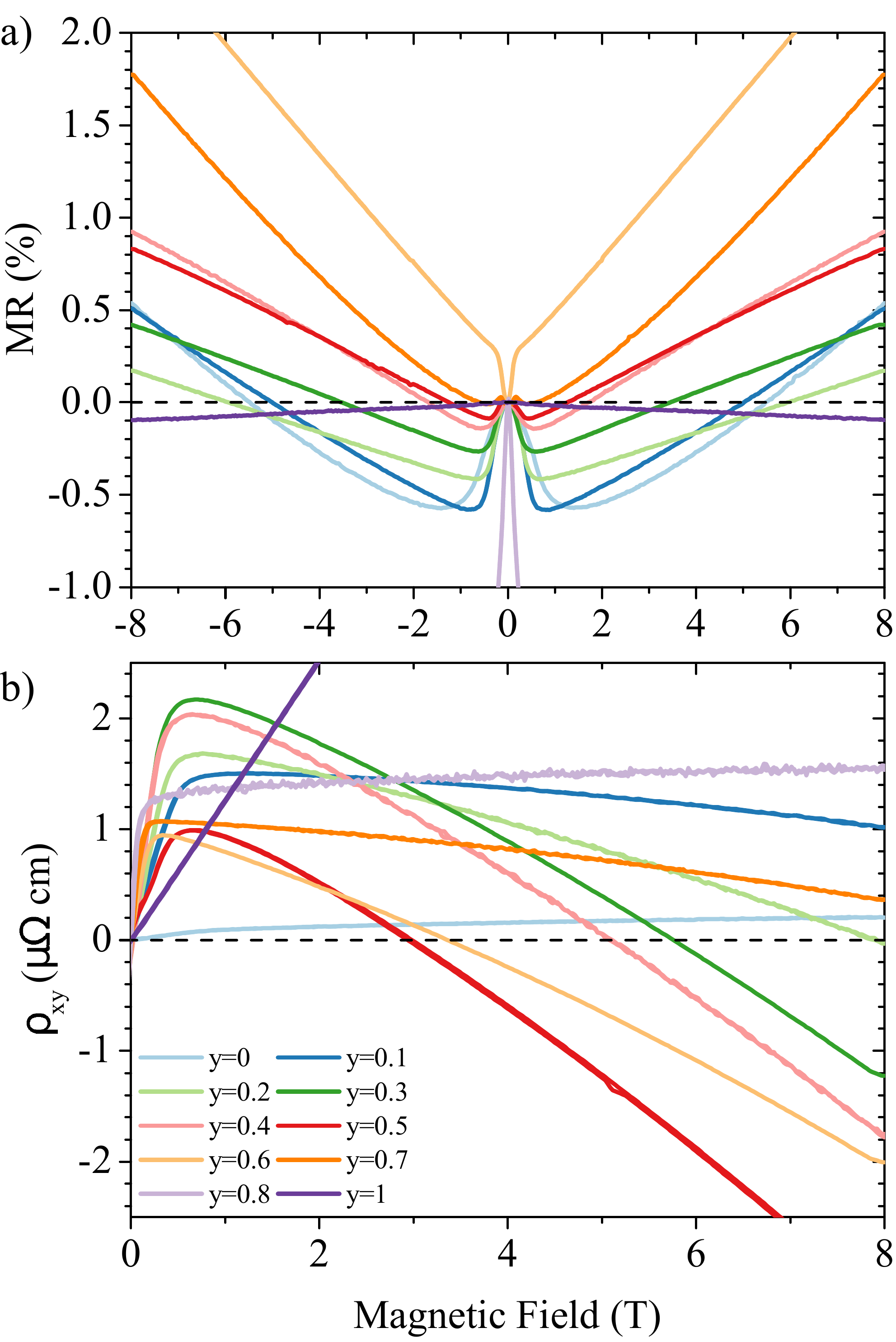}
\caption{(color online) Magnetotransport measurements with applied field out-of-plane at 5~K for (a) magnetoresistance and (b) total Hall resistivity, as a function of field strength for all concentrations of $y$.\label{expMR_Hall}}
\end{figure}

Magnetoresistance (MR) measurements were taken using an applied field perpendicular to the sample plane. The data at 5~K up to $\pm 8$~T are shown for $y = 0$ to 1 in Fig.~\ref{expMR_Hall}(a). In this orientation the applied field is parallel to the helix axis and the low-field ($\lesssim 1$~T) acts to distort the helix into the conical phase. On the increasing field, magnetic saturation is reached at $H_\mathrm{c}$, and the sample becomes uniformly magnetized. Below $H_\mathrm{c}$ the change in resistivity is due to a GMR-like mechanism where there is rotation of the moments in the conical phase, whilst at higher fields a variety of different mechanisms are at play, depending on the temperature regime \cite{Porter2014}. It is noteworthy that the high-field MR takes on an unusual linear form for values of $y$ at and around 0.5. Such linear magnetoresistance in the Fe$_{1-x}$Co$_x$Si system have been associated with its half-metallicity \cite{Guevara2004,Onose2005,Sinha2014}.

\subsubsection{Anomalous Hall Effect}
\begin{figure}[t]
\includegraphics[width=8cm]{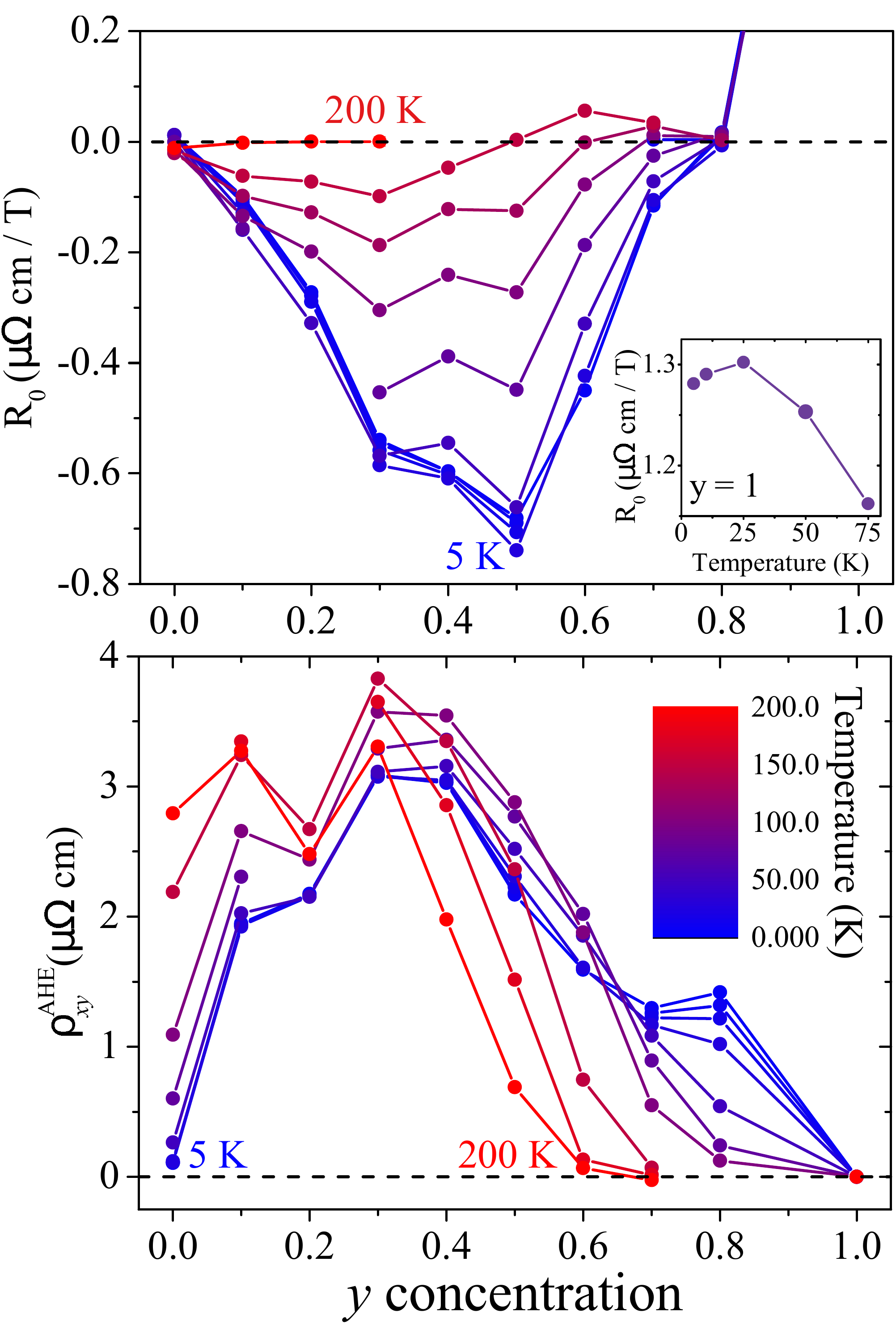}
\caption{(color online) Temperature dependence of the (a) ordinary Hall coefficient $R_0$ and (b) anomalous Hall resistivity $\rho_{xy}^\mathrm{AHE}$ for all concentrations of $y$ up to 200~K. Inset: $R_0$ for $y = 1$ as function of temperature. \label{expOHE_AHE}}
\end{figure}

The measured total Hall resistivity \pxy for each concentration of $y$ at 5~K is shown in Fig.~\ref{expMR_Hall}(b). For FeGe a positive OHE is seen with a small AHE at 5~K consistent with previous measurements\cite{Porter2014}. With the introduction of Co the OHE becomes negative and a sudden increase in the AHE is seen. As $y \rightarrow 0.5$ the OHE is found to increase to a maximum at $y = 0.5$ and with further addition of Co as $y \rightarrow 0.7$ the OHE decreases. At $y = 0.8$ the OHE becomes positive and for CoGe a large positive OHE is seen with no sign of any other contributions to the Hall effect. The OHE coefficient $R_0$, determined from the high-field Hall slope, is shown for all concentrations up to 200~K in Fig.~\ref{expOHE_AHE}(a).

One of the key features of ferromagnetic materials is the AHE, which arises due the spontaneous breaking of time-reversal symmetry and the presence of spin-orbit coupling. This cause electrons to undergo a spin-dependent transverse acceleration in the absence of an external magnetic field. Spin-polarization then leads to a charge imbalance and a measurable Hall voltage. The anomalous Hall resistivity can be extracted from the measurement of the total Hall resistivity by extrapolating the high-field (saturated) data to zero applied field in order to take into account the presence of the OHE. In Fig.~\ref{expOHE_AHE}(b) the AHE resistivity $\rho_{xy}^\mathrm{AHE}$ is shown for all concentrations $y$ as a function of temperature up to 200~K.

The AHE is known to have three contributions that arise from three separate mechanisms \cite{Nagaosa2010}. In the state where $M_z$ is saturated these can be decomposed as:
\begin{equation}
R_\mathrm{s} = \left[\alpha+\beta\rho_{xx}+b\rho_{xx}\right]\rho_{xx}.
\label{AHE}
\end{equation}
The first two terms, with pre-factors $\alpha$ and $\beta$, are the extrinsic scattering contributions to the AHE, termed the skew and side-jump scattering, respectively. The skew scattering arises due to an asymmetry in scattering rates for each spin that results in a net current traverse to the applied field and dominates in the \textit{very clean} metal regime. Meanwhile the side-jump scattering, which is impurity density independent, is due to a transverse shift of the electron trajectory upon scattering at the center with SOC, and becomes relevant in the \textit{moderately dirty} metal regime. Nevertheless, the dominant contribution in the \textit{moderately dirty} regime is that arising from the intrinsic mechanism $b\rho_{xx}^2$, which is due to the $k$-dependent topology of the electronic band structure arising from effective magnetic monopoles in momentum space.

\begin{figure}[t]
\includegraphics[width=8cm]{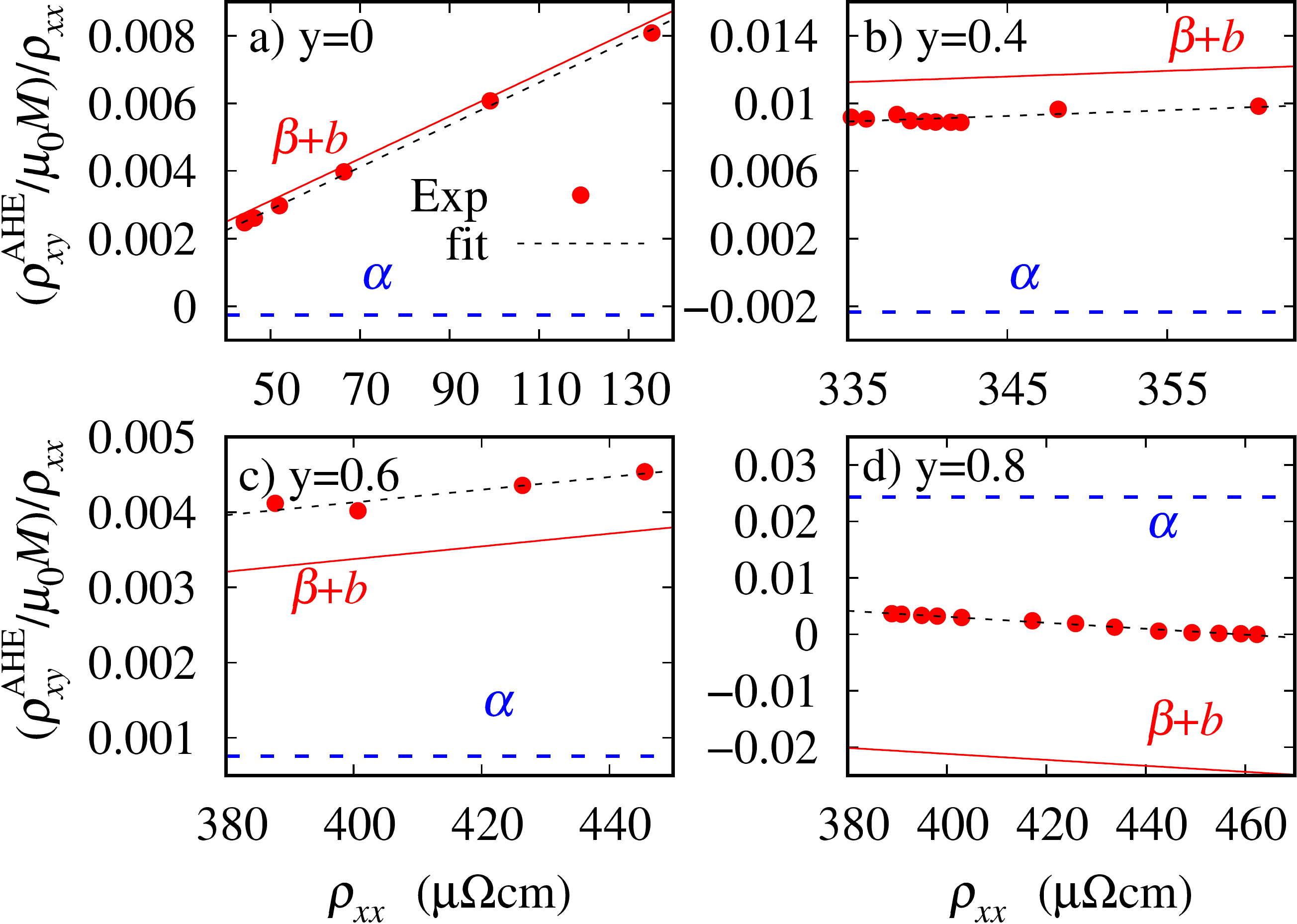}
\caption{(color online)  Plots of experimental data (red circles) for $(\rho_{xy}^\mathrm{AHE}/\mu_0M)/\rho_{xx}$ vs. $\rho_{xx}$ for (a) FeGe, (b) Fe$_{0.6}$Co$_{0.4}$Ge, (c) Fe$_{0.4}$Co$_{0.6}$Ge, and (d) Fe$_{0.2}$Co$_{0.8}$Ge. Linear fits are shown by dashed black lines. The skew scattering contribution ($\alpha$) is shown as a dashed blue line, and the side-jump and intrinsic contribution ($\beta + b$) as a solid red line.\label{fitAHE}}
\end{figure}

The different dependences on $\rho_{xx}$ can be used to separate out the skew scattering contribution ($\propto \alpha$) by fitting a straight line with the form
\begin{equation}
\left( \frac{\rho_{xy}^\mathrm{AHE}/\mu_0 M_z}{\rho_{xx}} \right) = \alpha + (\beta + b)\rho_{xx} \label{hallscale}
\end{equation}
to the data and determining its intercept, with $M_z$ taking its saturated value. The slope $(\beta + b)$ corresponds to a  combination of the side-jump ($\propto \beta$) and intrinsic ($\propto b$) contributions. In Fig.~\ref{fitAHE} this separation of the scattering density-dependent and -independent terms is shown for selected \FeCoGe concentrations $y$ by plotting the relationship between these two quantities as $T$ is varied and fitting Eq.~\ref{hallscale}. In the case of FeGe, the anomalous Hall effect is dominated by scattering density-independent mechanisms, i.e. those determined by $\beta$ and $b$, which agrees with previous work \cite{Porter2014}. However, for other concentrations the skew scattering term changes sign and magnitude as a function of concentration $y$. 	
	
\subsubsection{Topological Hall Effect} \label{Mag:THE}

It is well-known that skyrmions can be stabilized in helimagnetic systems by an external magnetic field. These skyrmions produce an effective magnetic field $B_\mathrm{eff}$ in real space that results in the topological Hall effect. To find the THE contribution the OHE and AHE can be found at high-fields (>~1~T), where any topological structure is destroyed by magnetic saturation, and subtracted from the total Hall effect, as per Eq.~\ref{halleffect} \cite{Neubauer2009,Kanazawa2011,Huang2012,Porter2014,Gallagher2017}. This leaves only the topological Hall contribution given by
\begin{equation} \label{eq:expTHE}
\rho_{xy}^\mathrm{THE}(H)=\rho_{xy}(H)-\left(R_0\mu_0 H + R_\mathrm{s} \mu_0 M_z(H) \right),
\end{equation}
where, as per Eq.~\ref{AHE}, $R_\mathrm{s} = \left[\alpha + (\beta + b)\rho_{xx}\right]\rho_{xx}$. For FeGe it has been previously shown that the skew-scattering term, $\alpha$, is negligible \cite{Huang2012,Porter2014} which is also seen here in Fig.~\ref{fitAHE} and so for simplicity a form of $R_\mathrm{s} = b\rho_{xx}^2$ is used where both the side-jump and intrinsic scattering contributions are combined into one parameter $b$ as they cannot be determined separately. By comparing $\rho_{xy}(H)/ \mu_0 H$ against $(b\rho_{xx}^2(H) M_z(H))/H$ above saturation, with $M_z(H)$ taken from the data in Fig.~\ref{expMH} and $\rho_{xx}(H)$ from the data in Fig.~\ref{expMR_Hall}, a linear fit is used to determine the parameters $R_0$ and $R_\mathrm{S}$ for the OHE and AHE contributions above saturation. For \FeCoGe, although we see an increase of $\alpha$ with increasing $y$, the inclusion of an $\alpha$ term is found to be negligible and the form of $R_\mathrm{s} = b\rho_{xx}^2$ can be used across the group without introducing any significant error.

\begin{figure*}[t]
\includegraphics[width=17cm]{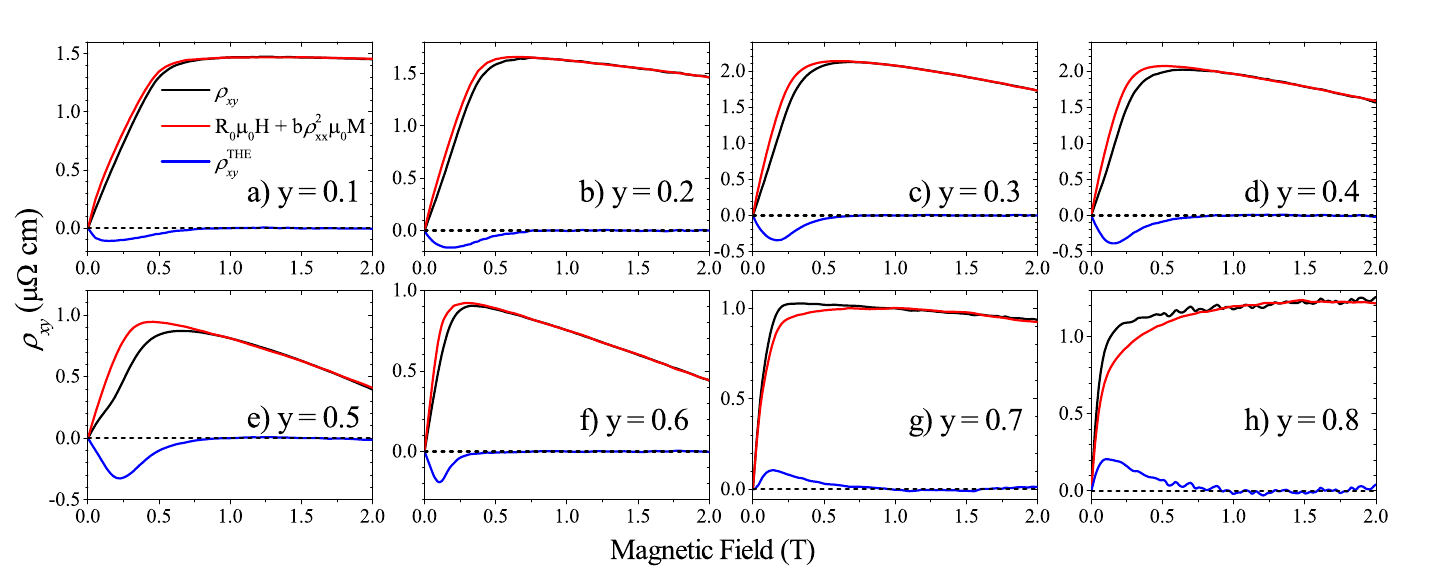}
\caption{(color online) Measured Hall resistivity $\rho_{xy}$, fitted data $R_0 \mu_0 H + b \rho_{xx}^2 \mu_0 M_z$ and the topological Hall resistivity $\rho_{xy}^\mathrm{THE}$ as a function of magnetic field at 5~K, for \FeCoGe with concentrations $0.1 \leq y \leq 0.8$ shown in (a-h). $\rho_{xy}^\mathrm{THE}$ results as a subtraction of fitted data from $\rho_{xy}$.}
\label{expTHE}
\end{figure*}

In Fig.~\ref{expTHE}, the $\rho_{xy}^\mathrm{THE}(H)$ obtained by this method for the positive field quadrant is plotted for $0.1 \leq y \leq 0.8$, where the average $\rho_{xy}^\mathrm{THE}(H)$ between the increasing and decreasing field sweeps is shown. (A detailed account of the analysis of the data for $y=0$ FeGe is given in Appendix~\ref{App_B}.) In all cases, the signal is close to zero for fields larger than $\sim 1$~T, implying that any skyrmion structure is completely unwound for all concentrations of $y$. The saturation fields in these signals are broadly in agreement with those obtained from the magnetometry (Fig.~\ref{expMH}) and MR (Fig.~\ref{expMR_Hall}). At lower fields the usual shape is a broad excursion with its extremum occurring at $\mu_0 H \approx 0.1$-0.2~T. In most cases this is a negative-going peak, but it is positive-going for $y = 0.7$ and 0.8. The sign and magnitude of the THE signal for FeGe, negative going and a few tens of n$\Omega$cm, is consistent with previous studies \cite{Huang2012,Porter2014,Gallagher2017}.

\section{Theory}

This rich set of effects of adjusting the Co content $y$ of the compound on the magnetic and magnetotransport properties require interpretation through first-principles theory. We carried out density functional theory (DFT) calculations using the full potential linearized augmented plane wave method within the generalized gradient approximation (GGA) exchange correlation (XC) functional for bulk crystals of \FeCoGens. We calculated the magnetic properties of the compound materials using the virtual crystal approximation (VCA)\cite{Bellaiche2000} along with the Vegard's lattice constant from the experimental lattice constants of pure CoGe and FeGe. The VCA averages the nuclear number of the Co and Fe ions with the weighting of the concentration of $y$. Our collinear calculations were converged with a plane wave cutoff of 4.0 a.u.$^{-1}$ and 512 {\bf k}-points in the full Brillouin zone as a starting point for spin-spiral calculations. We have found that the Perdew-Burke-Ernzerhof (PBE)\cite{Perdew1996} XC functional gives good results for collinear calculations and magnetic moments for the B20 compounds in the equilibrium XC-functional lattice constant. The PBE collinear calculations yielded magnetic moments of 1.2 and 0 $\mu_B$ in FeGe and CoGe respectively at the experimental lattice constant which can be compared with the experimental values of 1.0 and 0.0 $\mu_B$ respectively (see Fig~\ref{expMs}). The hybridization of the transition metal $d$-orbitals with the Ge $p$-orbitals induces a small moment of less than 6 \% on the Ge atom. We used the collinear calculations as a starting point to converge the spin-spiral calculations. Fig~\ref{expMs} shows the effect of using the VCA approximation with the Vegard lattice constant on the magnetic moment of the transition metal ion at different concentrations of $y$. The trend is correct but this method of calculation tends to overestimate the moment by a small amount with respect to our experimental results. This is due to the GGA correction to the local density approximation, which causes a large moment due to an underestimation of the atomic binding.

In the absence of spin-orbit coupling, the homogeneous spin-spiral systems can be described using the generalized Bloch theorem, without the expense of large super cell calculations. Homogeneous spin-spirals with a specific direction can be described by a reciprocal lattice vector {\bf q} with the rotation angle $\varphi$={\bf q$\cdot$R$_n$} at a given atomic site. The unit vector of the magnetization at each basis site is best described by
\begin{eqnarray}\label{2}
 {\bf \hat{m}}_n & = & \left( \sin\theta \cos(\varphi+ \tau_i){\bf\hat{e}_x} + \right. \nonumber \\
 & & \left. \sin\theta \sin(\varphi+ \tau_i){\bf\hat{e}_y} +\cos\theta{\bf\hat{e}_z} \right),
\end{eqnarray}
where $\theta$ is the cone angle and $\tau_i$ is the phase at site $i$. The exchange-correlation field {\bf B}$_\mathrm{xc}$ has the same form. When spin and real space are decoupled, all atoms of the spin spiral are equivalent. The angle between the local moment and the lattice differs at each atomic site leading to a generalization of Bloch theorem\cite{Sandratskii1998,Sandratskii1991,Sandratskii1986a,Sandratskii1991a}. This generalized Bloch theorem allows one to solve for the eigenstates of a variation of the Schr{\"o}dinger Hamiltonian where {\bf B}$_\mathrm{xc}$ rotates from one unit cell to the next. The eigenstates from this Hamiltonian take a similar form to the Bloch eigenstates, with an extra phase factor:
\begin{equation}\label{3}
\psi_{{\bf k},j}({\bf r,q})=\left(
   \begin{array}{r}
    e^{i({\bf k-q}/2)\cdot {\bf r}}\alpha_{{\bf k},j}({\bf r}) \\
    e^{i({\bf k+q}/2)\cdot {\bf r}}\beta_{{\bf k},j}({\bf r})
   \end{array}
   \right).
\end{equation}
Here $\alpha$({\bf r}) and $\beta$({\bf r}) are the periodic spinor functions. These Bloch functions become very useful in treating spin-spiral states \textit{ab initio}, avoiding expensive super-cell calculations.

If the magnetization rotates along a high-symmetry line for a homogeneous spin-spiral, one can adopt a quasi one-dimensional model\cite{Heide2009a} in which the energy is only a function of one variable, the spatial period length $\Lambda_\mathrm{h} = 2 \pi/ |{\bf q}|$. In this micromagnetic model, the discrete spins of a classical Heisenberg-like Hamiltonian are mapped to the continuum limit where the magnetization changes smoothly with a given parameter. When the magnetization vector field is constant in magnitude, the total energy is only dependent on the parameter \textbf{q}:
\begin{equation}\label{4}
 E({\bf q}) = A{\bf q}^2 + D{\bf q} + \bar K.
\end{equation}
From the micromagnetic model, $\bar K$ is the magnetocrystalline anisotropy (MCA) tensor, which we neglect for the cubic B20 compounds. Here $A$ is the spin-stiffness parameter that stems from the isotropic and non-relativistic exchange parameter $J_{ij}$ from the Heisenberg model. Parameter $D$ stands for the strength of the DMI, whose contribution to the energy is linear and anti-symmetric around ${\bf q}=0$ with respect to $D$.

We converged the spin-spiral calculations starting from collinear calculations to 24$^3$ {\bf k}-points in the full Brillouin zone for small values of the reciprocal spin-spiral vector and including the ferromagnetic state ${\bf q}=0$ for the \FeCoGe systems. The spin-spiral vector ${\bf q}=(q,q,q)$ is chosen along the $[111]$ direction and due to symmetry the DMI vector points parallel or anti-parallel to the chosen {\bf q}-vector. Results are similar for \textbf{q} in the $[001]$ and $[110]$ directions. We use the micromagnetic model to approximate the symmetric and antisymmetric exchange parameters for the B20 compounds under consideration. Here it is assumed that the exchange parameters are independent of the magnetic moments at the atomic sites. We made use of the J\"ulich \textsc{fleur} code to calculate the ground state of \FeCoGe and to calculate the DMI, AHE, and THE in these materials. In the cubic B20 compounds the DMI and exchange is seen to be isotropic. We calculated long wavelength spin-spirals in the $(q,q,q)$ direction and applied perturbation theory to calculate the DMI.

\subsection{Dzyaloshinkii-Moriya Interaction}

The DMI favors a non-collinear state in which the interaction energy is minimized when the angle $\varphi$ between spins at neighboring sites $i$ and $j$ is $\pi$/2. In addition to spin-orbit coupling being necessary for the DMI, inversion symmetry must be broken in the real space environment of the lattice. When spin-orbit coupling is treated along with a spin-spiral configuration, the generalized Bloch theorem fails and the system cannot be solved within the chemical unit cell. Within DFT one can apply first-order perturbation theory\cite{Liechtenstein1987} with respect to spin-orbit interaction
\begin{equation}\label{5}
 H_\mathrm{so}=\sum_i\xi_i({\bf r}_i){\bf \boldsymbol\sigma \cdot L}_i,
\end{equation}
to the solutions of the {\bf q}-dependent Kohn-Sham orbitals. Here the spin-orbit operator must be transformed by a unitary spin-rotation matrix to satisfy the spin-spiral boundary conditions. We take advantage of the Andersen force theorem where the small perturbation from the spin-orbit operator causes a change in the single particle energies. The energy with respect to ${\bf q}=0$ at small {\bf q} is linear with the slope and determines the strength of the DMI.

The perturbation energy of the spin-spiral state is given by
\begin{equation}\label{6}
 \Delta \mathcal{E}_{{\bf k},n}({\bf q})=\langle\psi_{{\bf k},n}({\bf q,r})|H_\mathrm{so}|\psi_{{\bf k},n}({\bf q,r})\rangle
\end{equation}
at each {\bf q}, where $\psi_{k,n}({\bf q,r})$ is the unperturbed spin-spiral state without spin-orbit coupling. The expectation value of the expression in Eq.~\ref{6} goes to zero in the collinear state, and results in the calculation of the DMI for small finite values of {\bf q}. It can be shown that the solutions to Eq.~\ref{6} are antisymmetric in {\bf q} with respect to the ferromagnetic state ${\bf q}=0$.

The DMI energy can be calculated as the sum of the energy bands of the perturbed spin-spiral states,
\begin{equation}\label{7}
 E_\mathrm{DMI}({\bf q})=\sum_{{\bf k},n}\Delta \mathcal{E}_{{\bf k},n}({\bf q})f(\mathcal{E}_{{\bf k},n}({\bf q})),
\end{equation}
where $f(\mathcal{E}_{n})=[e^{(\mathcal{E}_n-\mathcal{E}_F)/k_B\mathrm{T}}+1]^{-1}$ is the Fermi-Dirac distribution function. The DMI for all these different compositions goes to a constant value of $-$1 meV{\AA} for large temperatures. The Fermi broadening smears the energy eigenvalues making different concentrations indistinguishable (see Figure~\ref{DMIresults}). As the Fermi broadening goes to zero the DMI for all concentrations approaches different constant value in each case. The largest value is seen for $y=0$ in the FeGe clean case. However, the FeGe seems to show two sign changes as function of the Fermi broadening, whereas only concentrations greater than $y=0.5$ show a sign change from positive to negative.

\subsection{Hall Effects}

\subsubsection{Anomalous Hall Effect}
For the calculation of the AHE we used 64 Wannier functions with spin-orbit coupling for each unit cell. With the $s$-states lying far below ($\sim1-3$~eV) the occupied $p$-states, we neglect any contribution from the $s$-states. The calculations of the anomalous Hall conductivities are converged on a $512^3$ {\bf k}-point interpolated grid in the full Brillouin zone.

The calculation of the intrinsic contribution to the anomalous Hall conductivity (AHC) is straightforward using the Kubo formalism with Wannier function interpolation:
\begin{equation}\label{8}
 \sigma_{ij}  \overset{\Gamma \rightarrow 0 }{=}  \frac{2e^2\hbar}{\mathcal{N}} \sum_{{\bf k},n}^\mathrm{occ}\sum_{m\neq n}\mathrm{Im}\left[\frac{\mel{\psi_{{\bf k},n}}{v_i}{\psi_{{\bf k},m}}\mel{\psi_{{\bf k},m}}{v_j}{\psi_{{\bf k},n}}}{(\mathcal{E}_{{\bf k},m}-\mathcal{E}_{{\bf k},n})^2}\right].
\end{equation}
We introduce a disorder parameter to account for degenerate energy crossings in momentum space. This disorder parameter is well converged at 0.1 meV, for all VCA concentrations. In the absence of impurity scattering the intrinsic AHE mechanism originates in the momentum-space Berry curvature which can be determined based solely on the electronic structure of the pure crystal.  Eq.~\ref{8} is the intrinsic contribution to the AHE in the $\Gamma \rightarrow 0$ limit where a finite broadening $\Gamma$ is added to $(\mathcal{E}_{{\bf k},m}-\mathcal{E}_{{\bf k},n})^2$ for convergence. However, one can start from the Bastin formula\cite{Bastin1971,Crepieux2001b} in the eigenstate representation and the constant $\Gamma$ approximation for a more accurate description of the Kubo formula\cite{Freimuth2014a}:
\begin{equation}\label{9}
\begin{split}
 \sigma_{ij} = & \frac{e^2\hbar}{2\mathcal{N}}\sum_{{\bf k},n\neq m}\mathrm{Im}[\mel{\psi_{{\bf k},n}}{v_i}{\psi_{{\bf k},m}}\mel{\psi_{{\bf k},m}}{v_j}{\psi_{{\bf k},n}}] \times \\
 & \Big\{ \frac{\Gamma(\mathcal{E}_{{\bf k},m}-\mathcal{E}_{{\bf k},n})}{[(\mathcal{E}_{F}-\mathcal{E}_{{\bf k},n})^2+\Gamma^2][(\mathcal{E}_{F}-\mathcal{E}_{{\bf k},m})^2+\Gamma^2]} + \\
 &\frac{2\Gamma}{[\mathcal{E}_{{\bf k},n}-\mathcal{E}_{{\bf k},m}][(\mathcal{E}_{F}-\mathcal{E}_{{\bf k},m})^2+\Gamma^2]} + \\
 &\frac{2}{[\mathcal{E}_{{\bf k},n}-\mathcal{E}_{{\bf k},m}]^2}\mathrm{Imln}\frac{\mathcal{E}_{{\bf k},m}-\mathcal{E}_{F}-i\Gamma}{\mathcal{E}_{{\bf k},n}-\mathcal{E}_{F}-i\Gamma} \Big\}.
\end{split}
\end{equation}

Eq.~\ref{8} can be recovered from eq.~\ref{9} in the $\Gamma \rightarrow0$ limit. The two equations completely agree in the $\Gamma \rightarrow 0$ limit. However, they tend to diverge for finite broadening with different signs and magnitudes that differ by more than 50 \%. It can also be seen that the AHC is suppressed for large disorder $> 0.4$~eV. In the large $\Gamma$ regime, the AHC becomes positive for all concentrations but approaches zero as $\Gamma \rightarrow \infty$.

\begin{figure}[t]
\includegraphics[width=8cm]{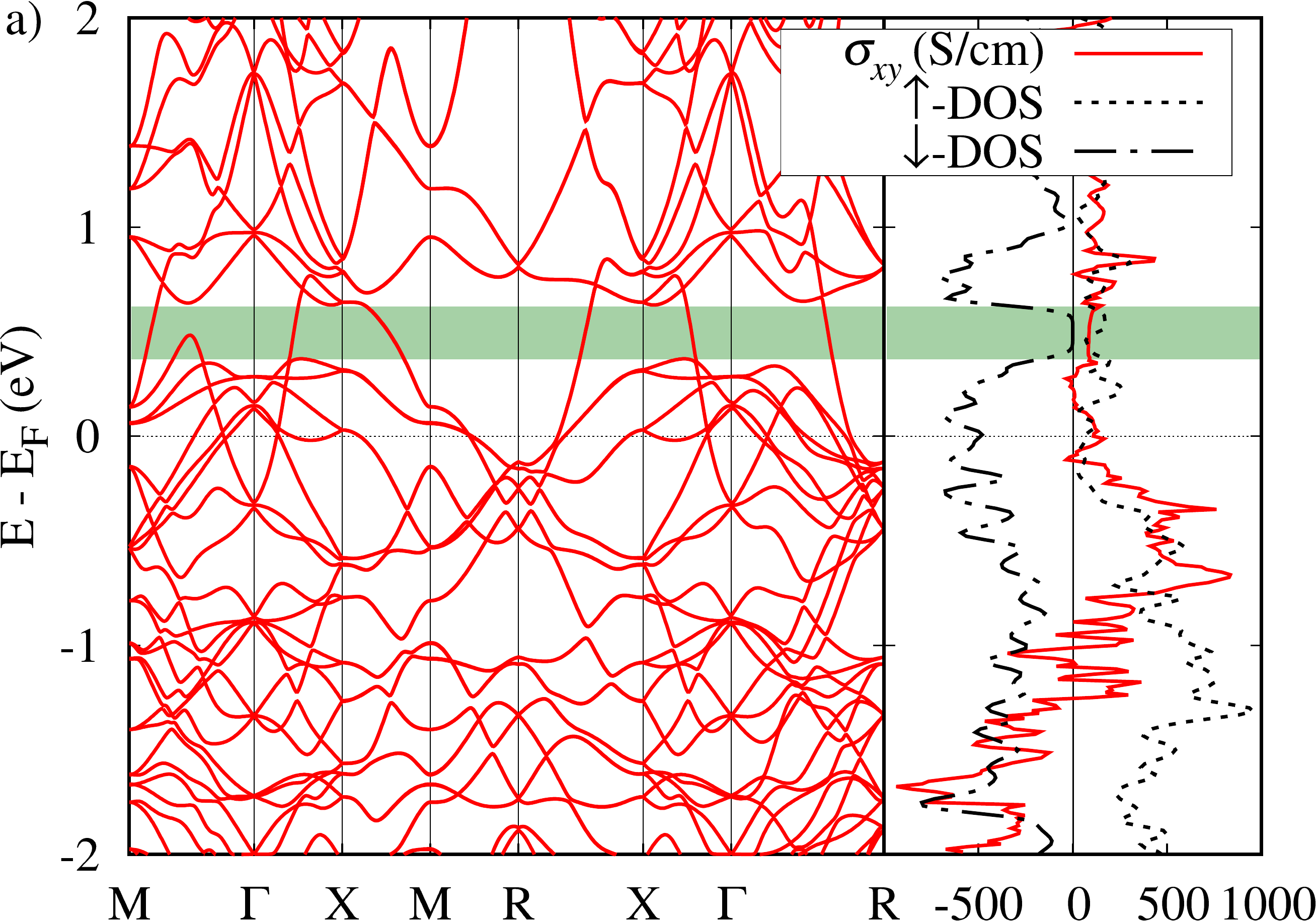}
\includegraphics[width=8cm]{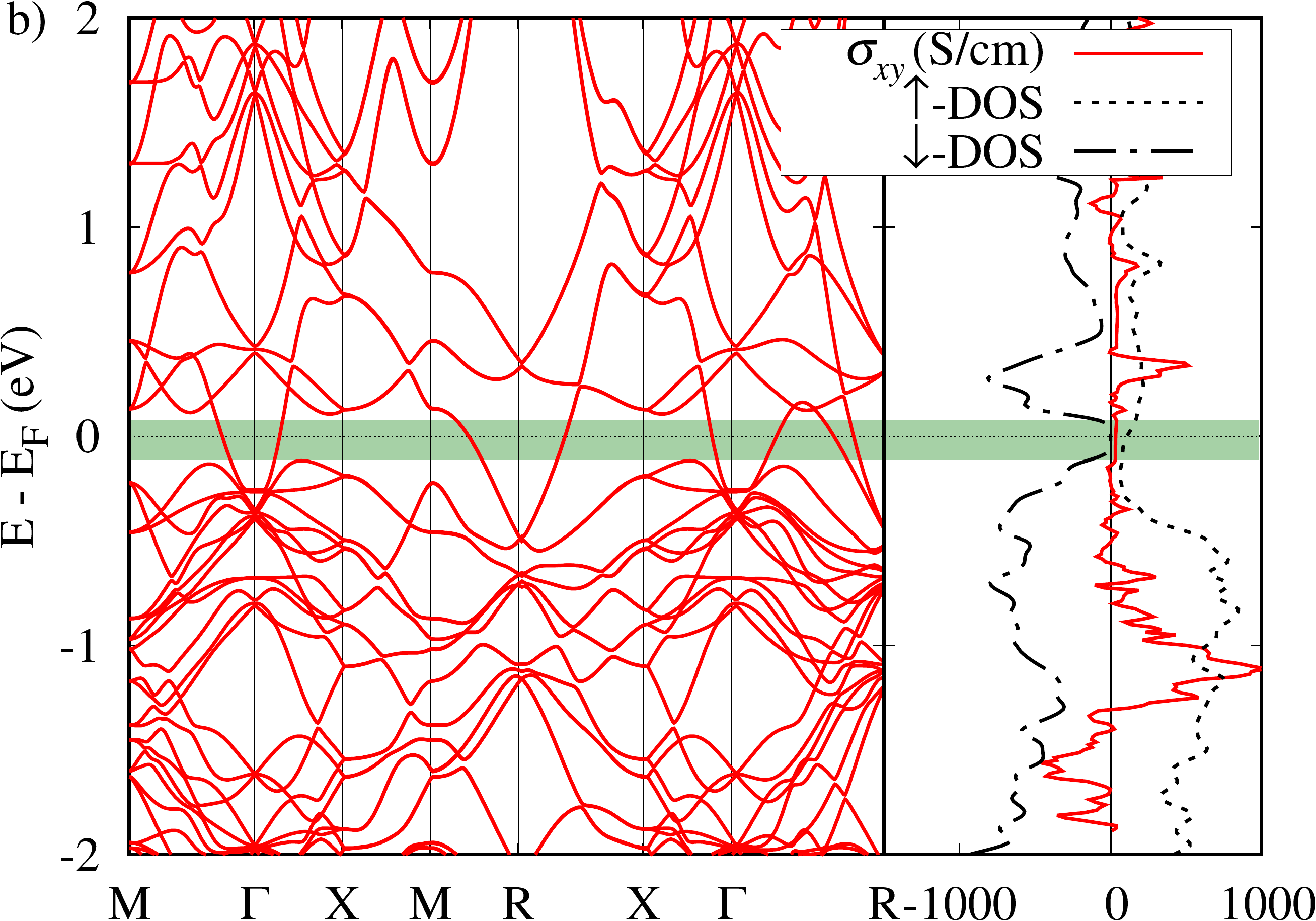}
\caption{(color online) Band structure of a) FeGe and b) Fe$_{0.4}$Co$_{0.6}$Ge with spin-orbit coupling. Dispersion relations are shown on the left. On the right the intrinsic anomalous Hall conductivity from first principles as a function of the Fermi energy is shown in red. In addition the black lines show the minority ($\uparrow$) and negative majority ($\downarrow$) density of states in black on an arbitrary scale. The green bar shows the range of energies in which the half-metallic state is present where there are no minority states. In the case of b) the Fermi level lies within this range. \label{bandFeGe}}
\end{figure}

In Fig.~\ref{bandFeGe} a) and Fig.~\ref{bandFeGe} b), we show the calculated band structures of FeGe and Fe$_{0.4}$Co$_{0.6}$Ge, respectively. Spin-orbit coupling was included in the calculation. The right panel for each of the figures shows the energy dependent intrinsic AHE is plotted in red, along with the spin-projected density of states (DOS), where the majority spin is plotted for positive values and the minority spin for negative values. The green bar in both figures shows the case where the minority DOS goes to zero, effectively showing a half-metal behavior. In FeGe, the half metallicity occurs around 0.5~eV above the Fermi level and so plays little role in the experimental properties. On the other hand, in Fe$_{0.4}$Co$_{0.6}$Ge this minority spin gap spans the Fermi level, consistent with the linear magnetoresistance observed experimentally for values of $y$ close to this one. This change in the band structure is due to the shift of the minority states caused by VCA approximation, and is only seen in our calculations at $y = 0.6$. In all cases the AHE calculated within the VCA approximation is constant in this gap. In this regime spin-flip transitions play an important role for ladder transitions across the Fermi energy, which may also influence the THE \cite{Zhang2011}. In experiments the AHE is more complex, where random substitution can cause a significant amount of disorder. Our CPA calculations of the electronic structure, shown in the appendix, reveal that the scattering for the majority states is strongly suppressed upon Co substitution, which in turn will also influence the AHE and even more so the THE.
	
\subsubsection{Topological Hall Effect}

For the calculation of the THE we used 32 Wannier functions in a collinear system without spin-orbit coupling for each spin channel in the unit cell. Again we neglect any contribution from the low-lying s-states. To calculate the conductivity tensors we used the interpolation technique based on the Wannier functions \cite{Franz2014}, using 512$^3$ {\bf k}-points in the full Brillouin zone. We used a lifetime broadening of 1 meV which resulted in converged values for the THE.

The calculation of THE requires semiclassical methods, where the approximation of the emergent magnetic field produced by the skyrmion lattice is valid for a slowly varying magnetization. In this limit the electron quasiparticles experience a Lorentz-like Hall force that is opposite in direction for each spin. We begin by calculating the ordinary Hall conductivities for each spin channel
\begin{flalign}\label{13}
\noindent
\sigma_{xy}^{\mathrm{OHE},s}(B^z) &=  -\frac{e^3B^z}{V\mathcal{N}}\sum_{\bf{k},ns}\tau_s^2\delta(\mathcal{E}_{F}-{\mathcal{E}_{{\bf k},ns}})\times \nonumber \\
& \left[(v_{{\bf k},ns}^x)^2m_{{\bf k},ns}^{yy}-v_{{\bf k},ns}^xv_{{\bf k},ns}^ym_{{\bf k},ns}^{xy}\right],
 \end{flalign}
along with the diagonal conductivities
\begin{equation}\label{12}
 \sigma_{xx}^{s}= \frac{e^2}{V\mathcal{N}}\sum_{\bf{k},ns}\tau_s\delta(\mathcal{E}_{F}-{\mathcal{E}_{{\bf k},ns}})(v_{{\bf k},ns}^x)^2,
 \end{equation}
where $\tau_s$, is the spin-resolved relaxation time, $V$ is the volume of the unit cell and $v_{{\bf k},ns}^i$ is the group velocity in the $i$ direction. We assume $\tau_s=\alpha g(\mathcal{E})_s^{-1}$ with $\alpha$ being constant and $g(\mathcal{E})$ being the spin dependent density of states, and with $m_{{\bf k},ns}^{ij}$ as the inverse effective mass tensor.

The THE constant $R_{yx}^\mathrm{THE}$ can be computed as the difference in the spin-resolved ordinary Hall conductivities divided by the square of the sum of the diagonal conductivities:
\begin{equation}\label{14}
 \rho_{yx}^{\mathrm{THE}}= R_{yx}^\mathrm{THE} {B_\mathrm{eff} } = \left[\frac{\sigma_{xy}^{\mathrm{OHE},\downarrow}-\sigma_{xy}^{\mathrm{OHE},\uparrow}} {(\sigma_{xx}^{\downarrow}+\sigma_{xx}^{\uparrow})^2}\right]{B_\mathrm{eff} }.
 \end{equation}
Here we assume the THE is due to a Lorentz-like force acting oppositely on differing spins with an emergent magnetic field $B_{\mathrm{ eff}}$.
Using this formalism, one is able to calculate the contribution that a nontrivial magnetic structure makes to the Hall effect from an electronic band structure (Eq.~\ref{14}). The topological constant that we calculate is independent of the relaxation time, since the OHE depends on the square of the relaxation time and the diagonal conductivities depend linearly on it.

\begin{figure*}[t]
\includegraphics[width=17.5cm]{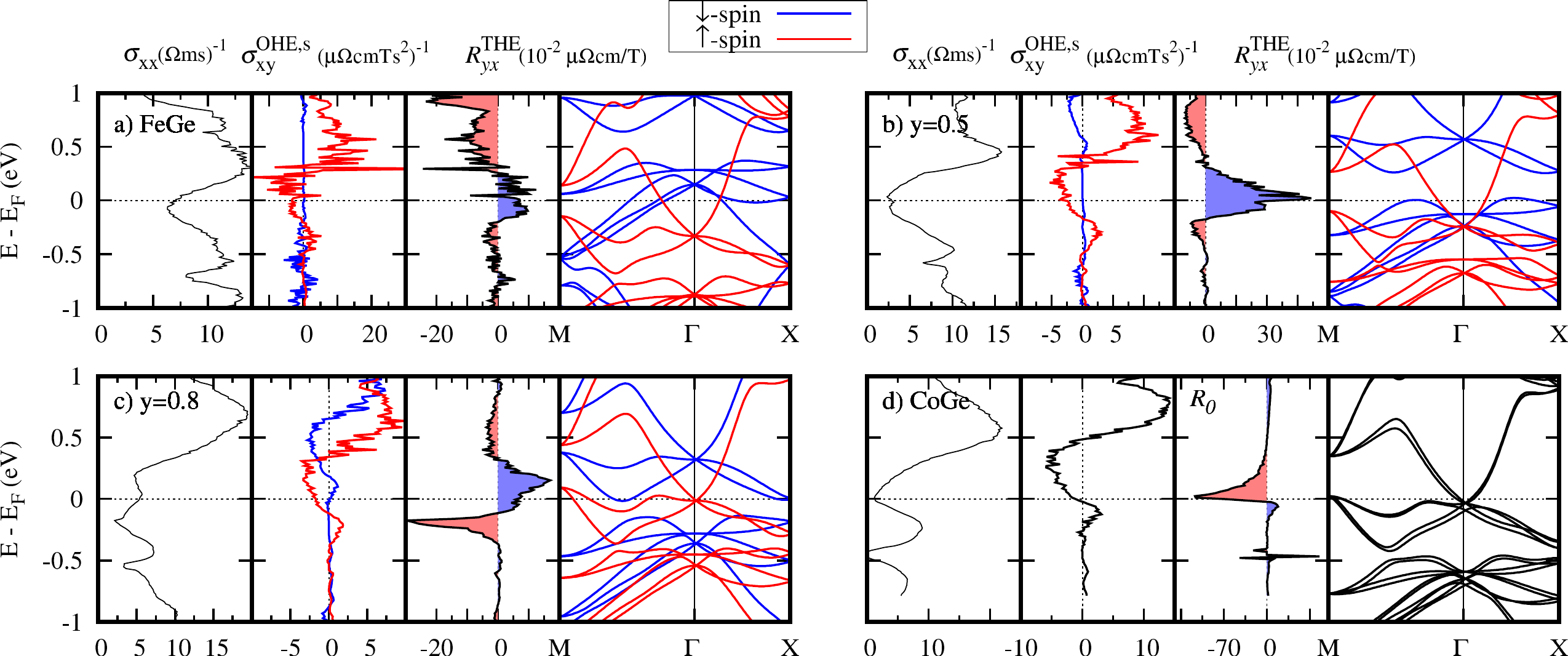}
\caption{(color online) Topological Hall effect in \FeCoGe for (a) $y=0$ FeGe, (b) $y = 0.5$, (c) $y = 0.8$, and (d) $y=1$ (CoGe). In each case the left-most panel shows first principles diagonal conductivity $\sigma_{xx}$ for the sum of both spin channels. The next panel shows the ordinary Hall conductivity for a given magnetic field $B^z$ spin minority ($\uparrow$) in red and majority ($\downarrow$) in blue. The second to last panel shows the first principles calculated topological Hall constant where blue shading is positive and red is negative. In the right-most panel the bands are plotted showing the band structure for each spin. For non-magnetic CoGe in panel (d), the effects are shown for the spin degenerate case.} \label{topooxx}
\end{figure*}

In Fig.~\ref{topooxx} we plot the topological Hall constant as a function of the Fermi energy in the third panel from the left for (a) FeGe, (b) $y=0.5$, (c) $y=0.8$, and the ordinary Hall constant for (d) CoGe. The figures also show the longitudinal conductivity, the spin resolved OHE conductivity, and spin resolved band structure in the left, middle left and right panels, respectively. In the spin-polarized cases, the sign of the topological Hall constant is determined by the OHE conductivity of $\uparrow$-spin states being an order of magnitude larger than the $\downarrow$-spin states. The magnitude of the topological Hall constant is determined by the low longitudinal conductivity. Therefore, the THE is maximized in the vicinity of the half-metallic gap for concentrations $0.4 \leq y \leq 0.6$. The ordinary Hall constant is maximized at 1.2~$\mu\Omega$cm/T (experimental value at $T = 5$~K is 1.3 $\mu\Omega$cm/T) in the nonmagnetic CoGe in the vicinity of the Fermi energy which may be due to the low conductivity. The addition of spin-orbit coupling shows no major change in the curves of OHE and THE. The trend of the theoretical results is congruent with that of the experimental OHE, which shows a peak around $y=0.4$ and the largest values for CoGe.

\section{Discussion}

\subsection{Helical Magnetism}
\begin{figure}[t]
\includegraphics[width=8cm]{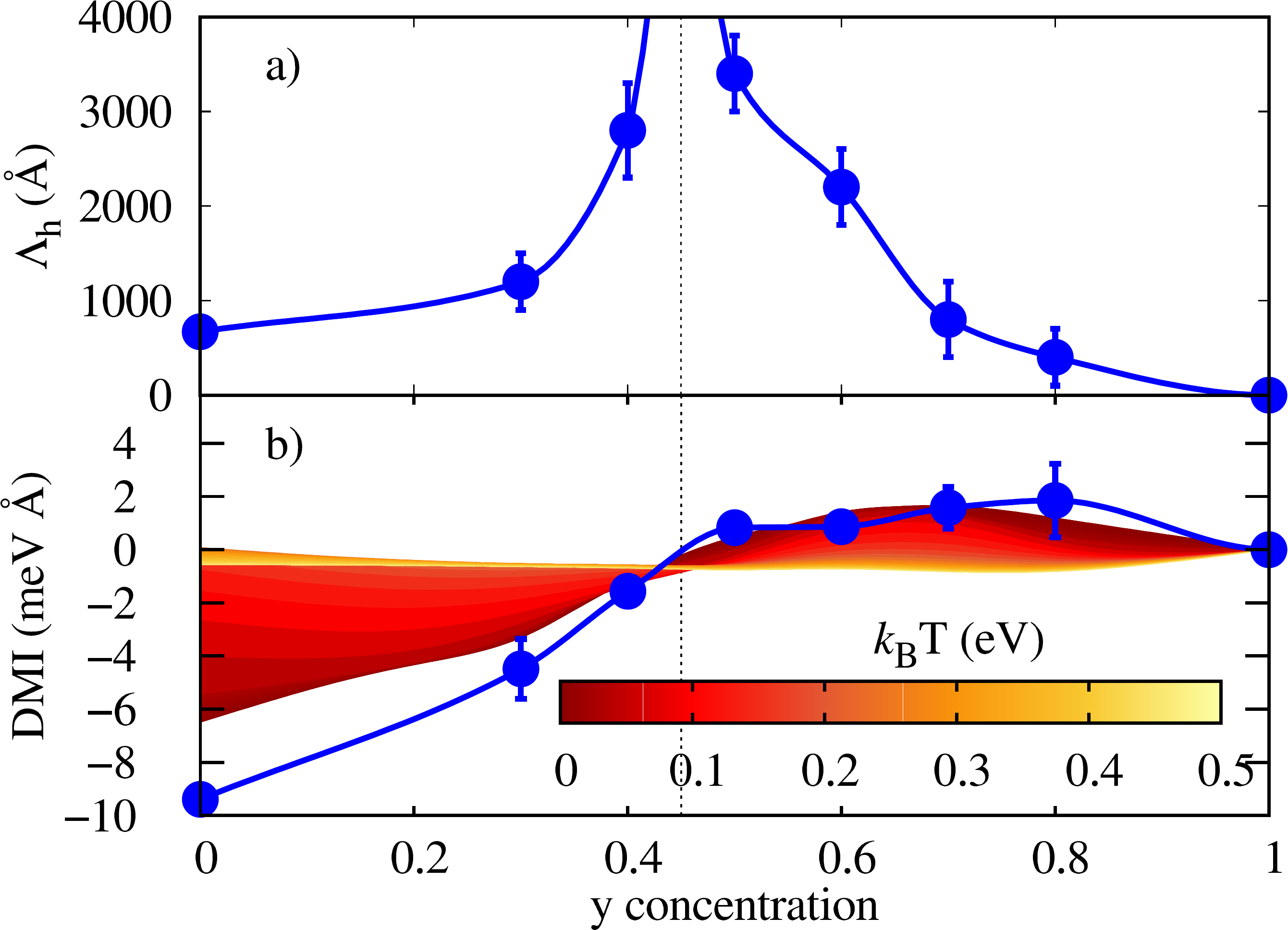}
\caption{(color online) Comparison of experimental and theoretical helical magnetic structures. (a) Experimental helical wavelength $\Lambda_\mathrm{h}$ as a function of concentration $y$, as determined by PNR. (b) Experimental DMI strength calculated from $D=(4 \pi A )/ \Lambda_\mathrm{h}$ in blue. The contour plot shows the DMI value that resulted from the first principles calculations, plotted as a function of concentration $y$ on the abscissa axis and Fermi broadening on the color axis. The vertical dotted line shows the position of the zero-crossing of the DMI strength. Blue solid lines are guides to the eye.}
\label{DMIresults}
\end{figure}

Let us now compare the results of the first-principles calculations with our experimental data. In Fig.~\ref{expPNR} we presented the PNR results from our \FeCoGe epilayers from which we can deduce the helical wavelength $\Lambda_\mathrm{h}$ for each sample at 50~K ($y = 0.6$ or less) and 5~K ($y = 0.7$ or greater), which is plotted in Fig.~\ref{DMIresults}(a). The wavelengths are the largest for intermediate values of $y$ in the region that the helix wavevector was found to go to zero in Ref.~\onlinecite{Grigoriev2014}.

It is then necessary to obtain values of the DMI strength in order to make a comparison with the first principles calculations of that quantity. To do so we make use of the fact that the helix wavelength is set by the ratio of exchange spin stiffness to DMI strength, written as $4 \pi \Lambda_\mathrm{h} = A/D$. We extracted the spin stiffness $A$ from our DFT calculations and used these to compute the values of DMI strength that are plotted as blue points in Fig.~\ref{DMIresults} (b). In these results we find the zero-crossing of the DMI to be at $y \sim 0.45$, slightly lower than zero-crossing of $y = 0.6$ found in bulk crystals \cite{Grigoriev2014}. Our experimental zero-crossing composition is reproduced by our DFT calculations, which is also at a slightly lower value than that calculated by the spin current method \cite{Kikuchi2016,Freimuth2017}. The experimental magnitude of the DMI is best reproduced in the calculation when the Fermi broadening is small.

\subsection{Hall Effects}

\begin{figure}[t]
\includegraphics[width=8cm]{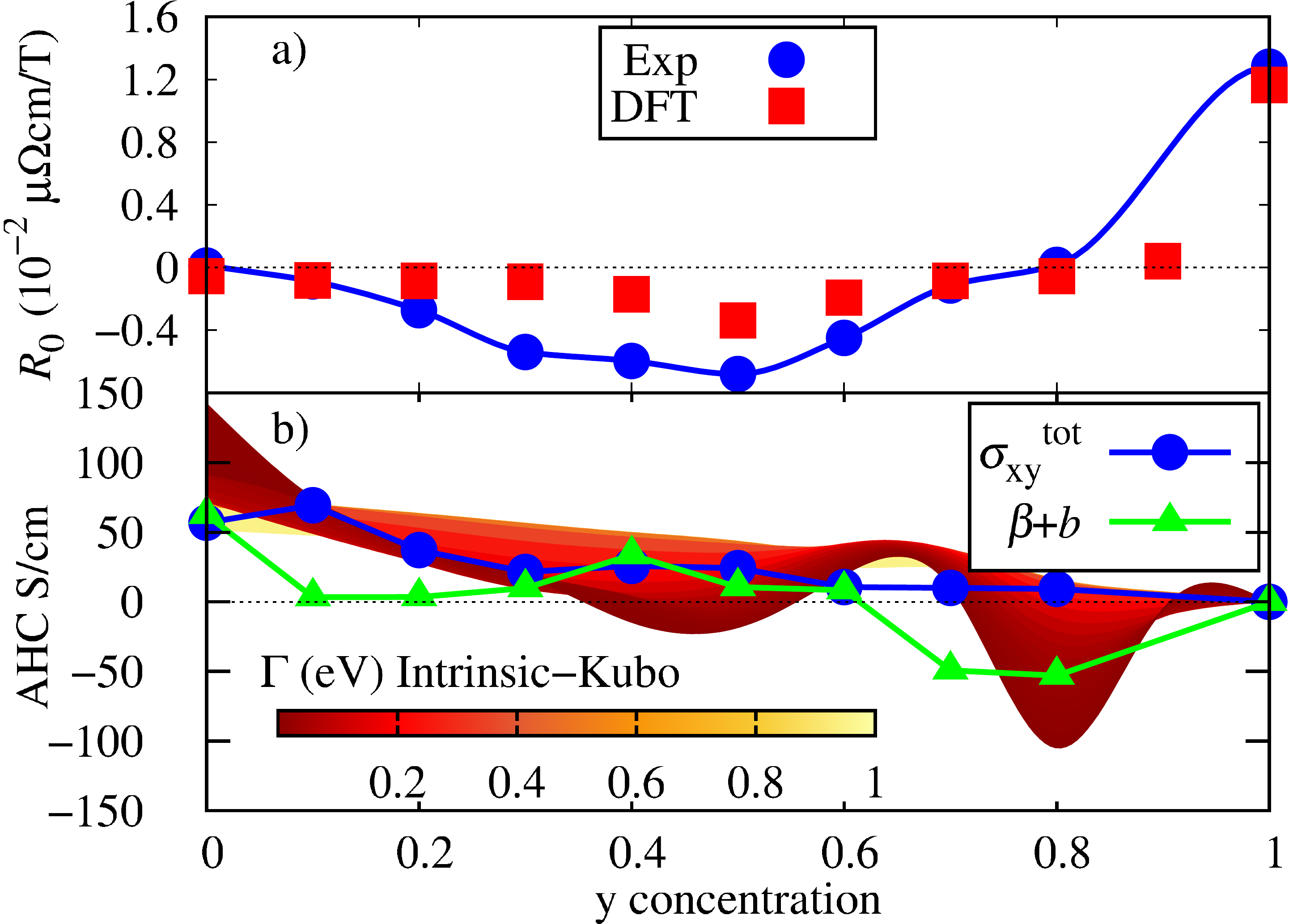}
\caption{(color online) Comparison of experimental and first principles Hall effects. (a) Experimental (circles) and theoretical (squares) ordinary Hall coefficients. (b) Experimental measurements and fits of the total anomalous Hall conductivity (blue dots) and the sum of the intrinsic and side-jump terms (green squares). The first principles calculation of the intrinsic anomalous Hall conductivity as a function of concentration and disorder is shown using the color axis. In both cases, experimental values were measured at 5~K. \label{AHCresults}}
\end{figure}

We now turn to our results for the Hall effects, starting with the ordinary Hall effect. In Fig.~\ref{AHCresults}(a) we plot the calculated values for the ordinary Hall coefficient $R_0$ as a function of $y$ alongside those measured at 5~K. There is good quantitative agreement between the two curves for the endmembers FeGe and CoGe. Both data sets also show a broad negative minimum centered around $y \sim 0.5$, and a change of sign as $y$ approaches 1: there is again agreement between experiment and theory in the zero-crossing at $y \sim 0.8$. Nevertheless, the experimental results are roughly 2-3 times larger than the DFT values in the central range of values of $y$.

Next we address the anomalous Hall effect. In Fig.~\ref{AHCresults}(b) we plot the experimental anomalous Hall conductivity $\sigma_{xy}^\mathrm{AHE}$ at 5~K as a function of $y$, determined from the Hall resistivities $\rho_{xy}^\mathrm{AHE}$ in Fig.~\ref{expOHE_AHE}(b) by inverting the resistivity tensor. Broadly, the trend is a gentle decline in $\sigma_{xy}^\mathrm{AHE}$ as $y$ increases.
In the fits in Fig.~\ref{fitAHE} we separated out the skew scattering contribution from those arising from side-jump scattering and the intrinsic Berry phase mechanism. In Fig.~\ref{AHCresults}(b) we also plot the sum of these two scattering density-independent parts of the AHC, as obtained from those fits. This has a more complex behavior, oscillating and changing sign as $y$ increases from zero before arriving at a value of zero for $y = 1$. Since we expect that these will be dominated by the intrinsic contribution, they can be compared to the results of the first principles calculations of the intrinsic AHC calculated through the Kubo formula in Eq.~\ref{9}. This also displays sign changes, particularly for low disorder, although they occur at somewhat different values of $y$ to those in the experimental data. The discrepancy in the intrinsic values can be attributed to the disorder that can not be captured by the VCA approximation, but is seen in the CPA calculations. The latter calculations also reveal that the spin-down states in the vicinity of the Fermi level are strongly localized by disorder, whereas the spin-up states remain almost Bloch-like. In this situation, the total conductivity of a material is defined solely by the spin-up states. This spin-selective localization leads to a high spin-polarization of the longitudinal electric current \cite{Chadov2013,Wollmann2015,Chadov2015}.

\begin{figure}[t]
\includegraphics[width=8cm]{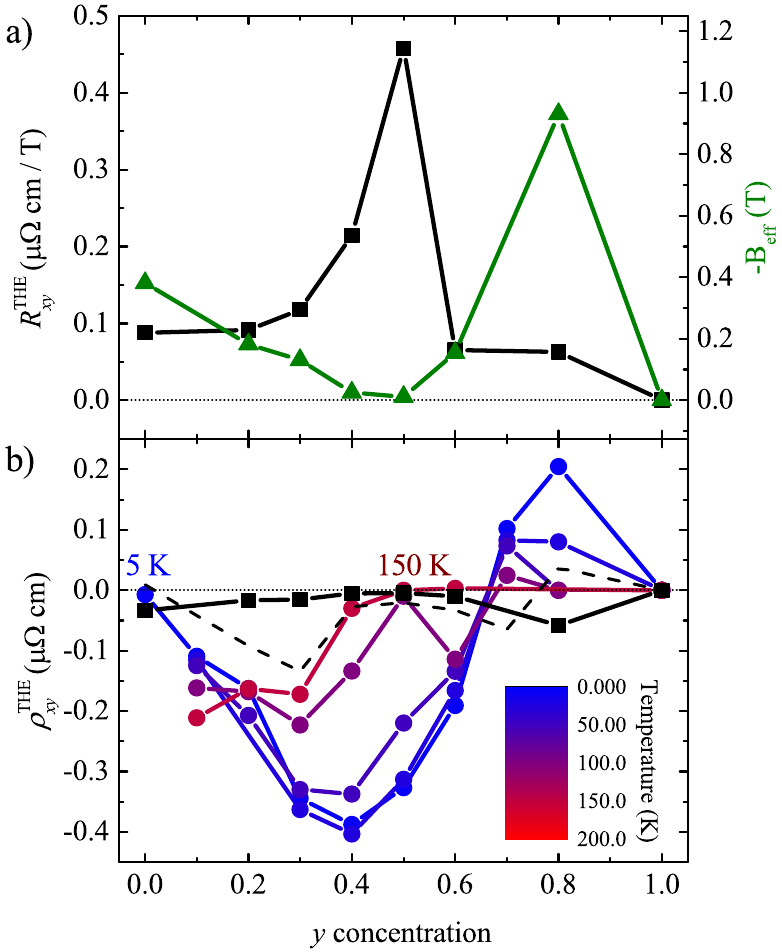}
\caption{(color online) (a) First principles calculations of the topological Hall constant $R_{yx}^\mathrm{THE}$ and emergent field $B_\mathrm{eff}$ as a function of concentration $y$. (b) Comparison of experimental topological Hall resistivity $\rho_{yx}^\mathrm{THE}$ (circles) from 5~K to 150~K against first principles calculation values (squares) using Eq.~\ref{14} as a function of concentration $y$. In both cases $B_\mathrm{eff}$ is estimated for a close-packed skyrmion lattice using the helix wavelength from the respective source. The dashed line shows the experimental estimated THE $\rho_{yx}^\mathrm{THE}\approx PR_0B_\mathrm{eff}$ at 5K with $P=-1$.}
\label{THCresults}
\end{figure}

Last, we discuss the topological Hall effect. We extracted the extremal values of $\rho_{xy}^\mathrm{THE}$ at 5~K from the data in Fig.~\ref{expTHE} and plot them as a function of $y$ in Fig.~\ref{THCresults}(b) at various temperatures (circles). The dashed line shows the estimated THE, $\rho_{yx}^\mathrm{THE}\approx PR_0B_\mathrm{eff}$, where $P=-1$ is assumed for the fully polarized state in the adiabatic limit \cite{Neubauer2009}. In Fig.~\ref{THCresults}(a) we plot the theoretical values of the topological Hall constant $R_{yx}^\mathrm{THE}$ (squares). There is a sharp peak at $y = 0.5$, reflecting the onset of the half-metallic nature of the theoretical band structure at higher composition.

Interconverting using the usual formula $R_{yx}^\mathrm{THE} = \rho_{xy} / B_\mathrm{eff}$ (Eq.~\ref{14}) for a direct comparison between theory and experiment is hampered by the fact that we do not know the exact value of the emergent field in our experiments. An upper limit on $B_\mathrm{eff}$ can be obtained by making use of the fact that the THE is proportional to the skyrmion winding number density, which for a fully dense triangular close packed lattice of skyrmions is determined only by the helix pitch. For skyrmions on a hexagonal lattice with helix pitch $\Lambda_\mathrm{h}$ one can estimate the emergent field $B_\mathrm{eff}= - \Phi_0 ( \frac{\sqrt{3}}{2\Lambda_\mathrm{h}^2})$, where the minus sign indicates an emergent field antiparallel to the external magnetic field and $\Phi_0$ is the magnetic flux quantum. The results of doing so are plotted in Fig.~\ref{THCresults}(a) (triangles), with very small emergent fields, as expected, near the zero-crossing of the DMI at $y \sim 0.45$. In our case the experimental emergent field is likely to be less than this ideal value, since the chiral grain structure seen in Fig.~\ref{chir} precludes skyrmion formation around the grain boundaries, preventing a fully dense skyrmion lattice from forming throughout the entire sample. The results of calculating the theoretically expected $\rho_{xy}^\mathrm{THE}$ from the two quantities plotted in Fig.~\ref{THCresults}(a) using Eq.~\ref{14} are shown in Fig.~\ref{THCresults}(b) (squares). The small values of $B_\mathrm{eff}$ near the DMI zero-crossing mean that the predictions for $\rho_{xy}^\mathrm{THE}$ in this range of $y$ are also small.

Nevertheless, the experimental data for $\rho_{xy}^\mathrm{THE}$ show a large negative peak at $y = 0.4$. However, the maximum size of the effective field in the regime is two orders of magnitude smaller than in FeGe, due to the very large helix pitch, implying a large increase in the $R_{yx}^\mathrm{THE}$ that is two orders of magnitude larger than the theoretical values. The reason for this discrepancy is unknown at the present time. Our experimental value for $\rho_{xy}^\mathrm{THE}$ for FeGe is consistent with those previously reported \cite{Huang2012,Porter2014,Gallagher2017}. Furthermore in the adiabatic limit \cite{Neubauer2009} the experimental values of $R_0$ and $B_\mathrm{eff}$ determine the maximal THE that follow the trend in magnitude of the theoretical calculations. Theory overestimates $\rho_{xy}^\mathrm{THE}$ for FeGe, which is expected given that in the theory we assume a fully dense skyrmion lattice that will not be present experimentally. Nevertheless, the theory dramatically underestimates $\rho_{xy}^\mathrm{THE}$ in the range of $y$ where experimentally we see the large negative extremum.  It is hard to imagine that the skyrmion lattice can be overdense by the two orders of magnitude required to explain this discrepancy. This disagreement seems even more striking given that in this regime of very large skyrmions and small emergent field the adiabatic picture used for theoretical estimates should be in principle more valid than for pure FeGe which exhibits skyrmions of smaller size and faster rotating magnetization.

Understanding the reasons behind this discrepancy presents a valid challenge, to be addressed in future studies. Given the overall good agreement between theory and experiment in describing the ordinary Hall effect around $y \approx 0.5$, it seems reasonable to disregard the influence of the exact details of the scattering processes as encoded in the $k$-dependence of the relaxation times, which was omitted in the present study by working in the constant relaxation time approximation, on the overall magnitude of the THE as seen in experiment. The fact that the discrepancy exists where the scale of the spin textures is expected to be large means that the adiabatic approximation should hold. Possible reasons for discrepancy could be thus related to following aspects: (i) the fact that the method of extracting $\rho_{xy}^\mathrm{THE}$ from the measured $\rho_{xy}$ expressed in Eq.~\ref{eq:expTHE} might, for some reason, not capture all relevant effects for intermediate values of $y$ in spite of reproducing prior work for $y=0$; (ii) prominence of additional mechanisms to the transverse Hall resistivity beyond the ones considered here for the AHE and THE, as expressed by Eq.~\ref{14}, that stem from the finite chirality of the magnetization and can be very sensitive to the details of disorder in the samples, see e.g. Ref.~\onlinecite{Ishizuka2018}; (iii) grain boundary scattering at the boundaries of the chiral grains playing some role when the helix pitch is large compared to the lateral grain size; or, (iv) the presence of exotic chiral spin textures in experiments which deviate significantly in their shape and properties from conventional skyrmions, which were taken as the foundation for the analysis of transport properties observed here. For example, there are a number of recent studies reporting the observation of so-called chiral bobbers\cite{zheng2018,ahmed2018} whose key feature is the presence of a singular Bloch point\cite{Rybakov2015}, and which could in principle exhibit transport properties radically different from those stemming from the adiabatic description.

\section{Conclusion}

In summary, we have carried out a comprehensive experimental study of the magnetic and transport properties of B20-ordered \FeCoGe epitaxial films for $0\leq y \leq 1$, complemented by DFT calculations.

The saturation magnetization was observed to decline gradually from the bulk-like value of about one Bohr magneton per Fe for FeGe ($y=0$) to zero for non-magnetic CoGe ($y=1$), reproduced in our calculations. The measured helix pitch diverges for $y \sim 0.45$, reproduced in the calculations as a zero-crossing of the DMI at that composition.

We found several unusual transport properties in our experiments on epilayers with intermediate values of $y$. These include peaks in the anomalous and topological Hall resistivity around $y \sim 0.5$. Our calculations suggest that these are associated with a high degree of spin-polarization in this regime. In particular, they predict a half-metallic state for $y = 0.6$, which is consistent with the observation of a linear magnetoresistance at high fields in the epilayer with that composition.

The most intriguing feature, which is still not completely understood is the THE, for which there is a large discrepancy between experiment and theory for intermediate values of $y$. In FeGe the THE shows reasonable agreement between experiment and theory, given that we assume a perfect skyrmion lattice in theory that is certain not to exist in our chiral-grain samples, and so we expect the experimental value of $\rho_{xy}^\mathrm{THE}$ to fall short of the theoretical upper limit by a factor of a few times. The very large experimental values of $\rho_{xy}^\mathrm{THE}$ around the peak at $y = 0.4$ far exceed the theoretical estimates. This discrepancy remains unexplained and presents a challenge for the future. Cutting edge magnetic tomography techniques are now being developed \cite{donnelly2017} that may be able to unravel the details of the spin textures in such layers. On the other hand, advances in theoretical methods better able to describe disorder such as scattering from chiral grain boundaries and inhomogeneous emergent fields can also be expected to provide a more accurate description of such systems.

Our results provide a comprehensive data set for the magnetic and magnetotransport properties of \FeCoGe epitaxial B20 films, and show that DFT calculations can provide a good description of how the magnetization and DMI vary with composition $y$. Our calculations also predict a half-metallic regime for $y \approx 0.6$, corroborated by the experimentally observed linear magnetoresistance. Nevertheless, our results require more intense investigations into the topological Hall effect in large non-trivial spin textures.

\begin{acknowledgments}
We would like to thank Christopher Morrison for assistance with the patterning of the Hall bars. This work was supported in part by the Science and Technology Facilities Council and the Hitachi Cambridge Laboratory. We are grateful to the ISIS Neutron and Muon Source for the provision of PNR beam time. This research was also supported by the Alexander von Humboldt Foundation, the ERC Synergy Grant SC2 (No. 610115), the Transregional Collaborative Research Center (SFB/TRR) 173 SPIN+X, and Grant Agency of the Czech Republic grant No. 14-37427G. Y. M. and F.F. acknowledge funding from the German Research Foundation (Deutsche Forschungsgemeinschaft, Grant No. MO 1731/5-1).  S.B. and Y.M. also acknowledge support by the Deutsche Forschungsgemeinschaft (DFG) through the Collaborative Research Center SFB 1238.”  C. H. M., S. B. and Y. M. acknowledge funding from the European Union's Horizon 2020 research and innovation programme under grant agreement number 665095 (FET-Open project MAGicSky).  We also gratefully acknowledge J\"ulich Supercomputing Centre and RWTH Aachen University for providing computational resources.

The data associated with this paper are openly available from the University of Leeds data repository at https://doi.org/10.5518/368/.
\end{acknowledgments}	

\clearpage

\appendix

\section{Characterization of \FeCoGe Epilayers} \label{App_A}

In Table~\ref{tab_s1} we display the results of measuring the lattice constant $a$ by XRD at the $(111)$ out-of-plane Bragg reflection, as well as the thickness of the \FeCoGe layer determined by XRR $t_\mathrm{XRR}$ and by PNR  $t_\mathrm{PNR}$ fitted using the GenX software \cite{Bjorck2007}. For $y = 0.3$ and 0.8 two samples were used in this study due to insufficient material left for magnetotransport sample fabrication. The substitute sample was produced using the same growth procedure and chosen due to the similar $a$ shown in Table~\ref{tab_s1} as well as similar \Ms and \Tc (not shown).

\begin{table}[ht]
\caption{Summary of values from XRD for \FeCoGe lattice constant $a$ and \FeCoGe layer thickness from XRR $t_\mathrm{XRR}$ and from PNR $t_\mathrm{PNR}$ measurements for compositions $0 \leq y \leq 1$ ($^\star$sample used for magnetometry and PNR measurements, $^\dag$sample used for magnetotransport measurements). \label{tab_s1}}
	
	\begin{ruledtabular}
		\begin{tabular}{llll}
			$y$  & $a$ (nm) & $t_\text{XRR}$ (nm)   & $t_\text{PNR}$ (nm)  \\   \hline
			0   & 0.4691(1)	& 67.8(2) 	      	& 67.7 \\
			0.1 & 0.4691(1) & 70.6(4)  	& -     \\
			0.2 & 0.4680(1)	& 76.6(8)	   	& -   \\
			0.3$^\star$ & 0.4675(1) & 64.1(1)     	& 63.6 \\
			0.3$\dag$ &	0.4676(1)	& 116(4) & - \\
			0.4 & 0.4670(1) & 63.3(4)	    	& 61.6  \\
			0.5 & 0.4662(1) & 65.9(2)	     	& 66.4 \\
			0.6 & 0.4655(1) & 63.6(3)	     	& 62.0 \\
			0.7 & 0.4649(1) & 61.8(4)	     	& 61.9 \\
			0.8$^\star$	& 0.4644(1)		& 64.9(2)	& 66.4 \\
			0.8$^\dag$ & 0.4645(1) & 62.2(2)	     	& - \\
			1   & 0.4630(1) & 63.3(4)	    	& - \\
		\end{tabular}
	\end{ruledtabular}
\end{table}

\section{Topological Hall Effect Data Analysis} \label{App_B}

\begin{figure}[t]
\includegraphics[width=8cm]{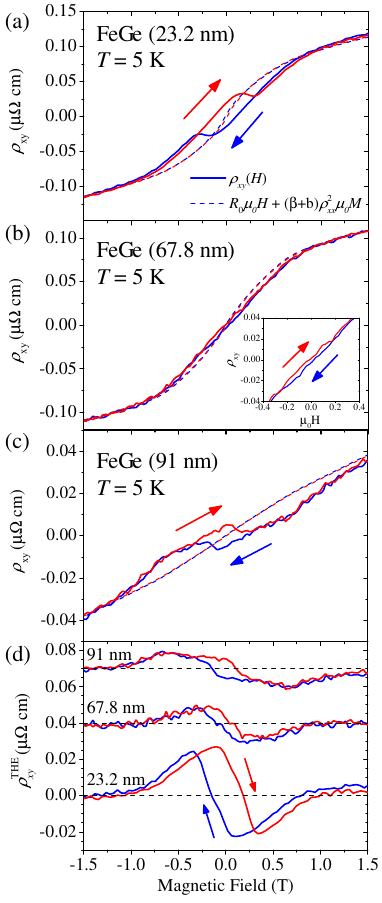}
\caption{(color online) Hall resistivity $\rho_{xy}$ measurements at 5~K for three FeGe films with film thickness (a) 23.2~nm, (b) 67.8~nm and (c)~91 nm shown with solid lines for increasing (red) and decreasing (blue) field sweeps. Dashed lines with corresponding color show the fitted $R_0 \mu_0 H + (\beta + b) \rho_{xx}^2 M_z$ data for each field sweep direction with the same color code. The inset in (b) shows a zoomed view of the low-field data. All three films show a significant deviation between the measured and fitted data at low-fields. (d) Resulting Topological Hall effect $\rho_{xy}^\mathrm{THE}$ contribution from subtraction of fitted data from measured data for each field.}\label{App_B:FeGe_pxy_5K}
\end{figure}

To verify the procedure used to extract the THE resistivity $\rho_{xy}^\mathrm{THE}$ the method was applied to samples of FeGe ($y =$ 0) with varying thickness. This method has been used in several reports on FeGe samples previously \cite{Huang2012,Porter2014,Gallagher2017}, so the results on our samples here have a basis for comparison. The measured Hall resistivity $\rho_{xy}(H)$ for three different thicknesses of FeGe films at 5~K are shown in Fig.~\ref{App_B:FeGe_pxy_5K}(a-c) alongside the scaled magnetometry data $R_0 \mu_0 H + (\beta + b) \rho_{xx}^2 \mu_0 M_z$ based on a high field fit to obtain $R_0$ and $R_\mathrm{s} = (\beta + b) \rho_{xx}^2$. It is worth noting that there is a hysteretic feature in the centre of the Hall effect loop that is not present in the magnetomtry data, indicating that there is some sort of additional contribution to the Hall resistivity at low fields. This is the topological Hall contribution.

For each thickness of FeGe, Fig.~\ref{App_B:FeGe_pxy_5K}(a-c) shows a deviation between the measured data (solid lines) and the predicted response based on the OHE and AHE alone (dashed lines). For the 23.2~nm and 91~nm films, $\rho_{xy}$ is non-monotonic at low fields (-0.5~T$<\mu_0 H<$0.5~T), where the gradient changes sharply from positive to negative (on the increasing field sweep) before returning, which is characteristic of a THE contribution.

The resulting $\rho_{xy}^\mathrm{THE}$ is found by subtracting the fitted OHE and AHE contributions from the measured data using Eq.~\ref{eq:expTHE}, and is shown in Fig.~\ref{App_B:FeGe_pxy_5K}(d). Each FeGe film shows a hysteresis effect that is the result of the hysteresis seen in the resistivity that is not present in the magnetization. The shape, sign, and order of magnitude of the THE resistivity around the extremum is consistent with previous literature on FeGe \cite{Huang2012,Porter2014,Gallagher2017}.

\section{First Principles Calculations} \label{App_C}

\begin{figure}[t]
\includegraphics[width=8cm]{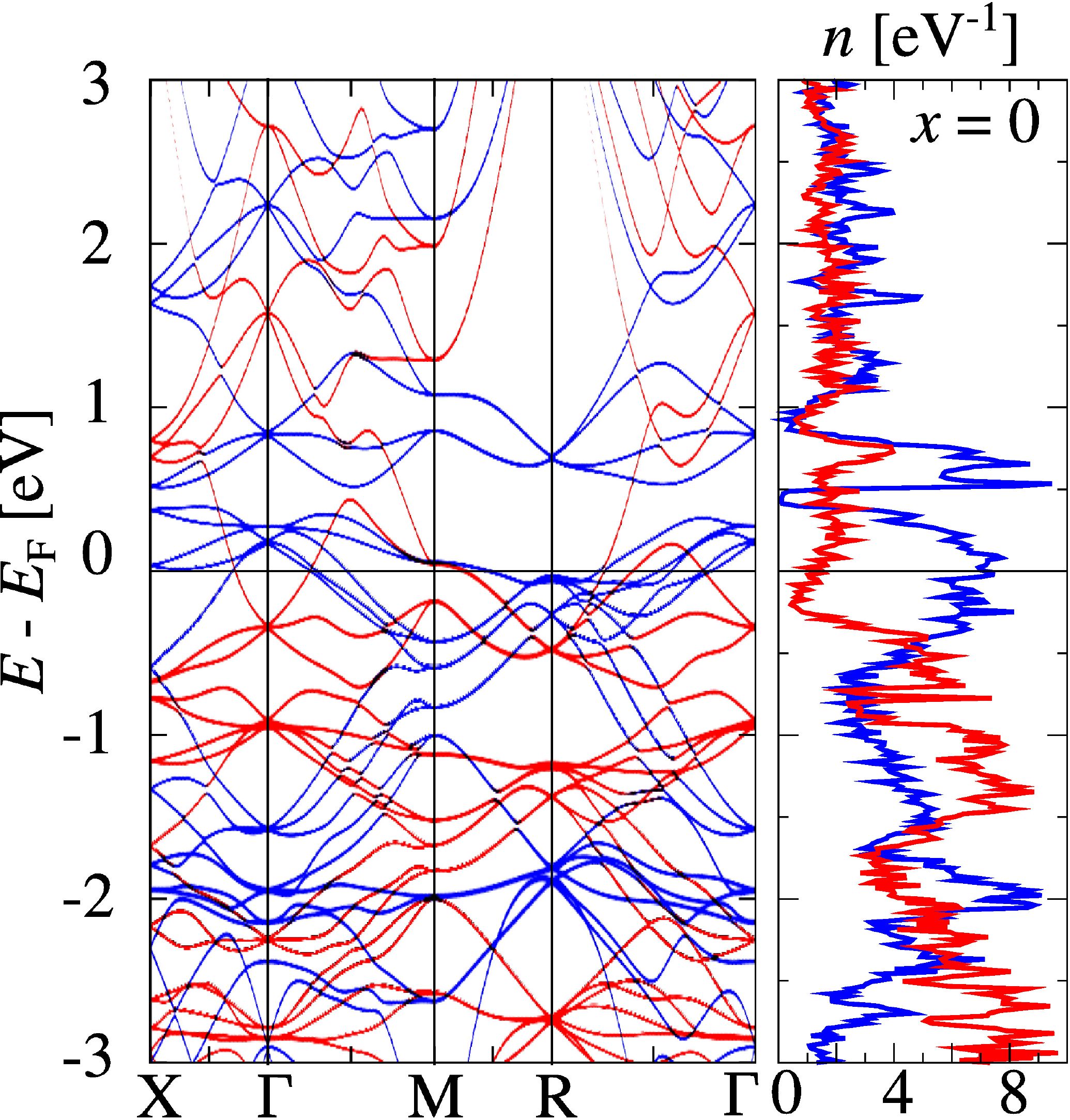}
\caption{(color online)  The left panel shows the band structure for the  spin resolved FeGe for $\uparrow$-spin in red and the $\downarrow$-spin in blue. The right panel shows the spin resolved DOS for the same color scheme as the bands.}
\label{CPA1}
\end{figure}

\begin{figure}[t]
\includegraphics[width=8cm]{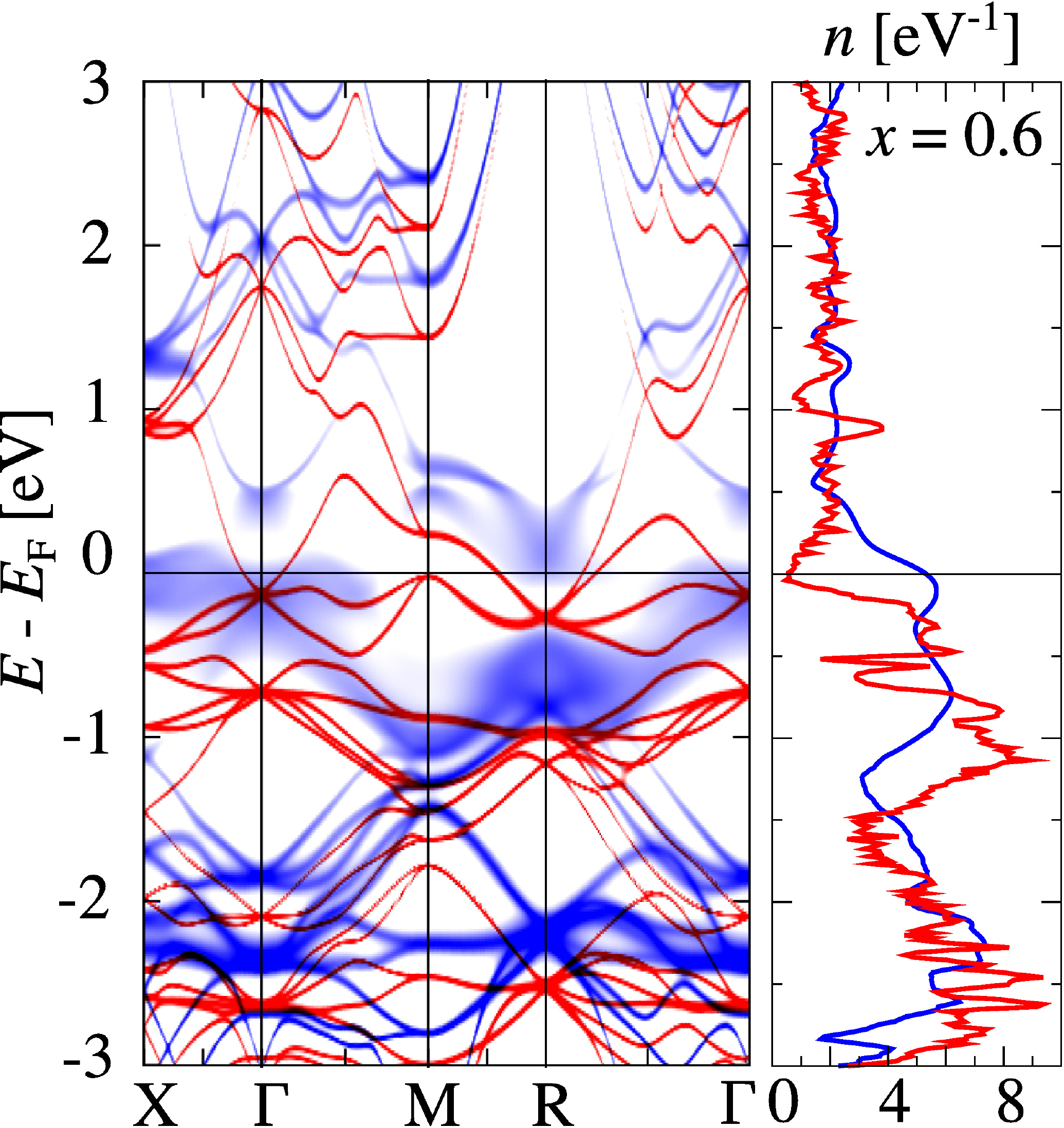}
\caption{(color online) The left panel shows the band structure for the  spin resolved Fe$_{0.4}$Co$_{0.6}$Ge for $\uparrow$-spin in red and the $\downarrow$-spin in blue. The right panel shows the spin resolved DOS for the same color scheme as the bands.}
\label{CPA2}
\end{figure}

The results of first principles calculations of the band structures and DOS for FeGe using the SP-KKR CPA method are shown in Fig.~\ref{CPA1}, and the equivalent results for the half-metallic Fe$_{0.4}$Co$_{0.6}$Ge are shown in Fig.~\ref{CPA2}. Here we see that the in the half-metallic Fe$_{0.4}$Co$_{0.6}$Ge that there is strong spin disorder on the spin down states which penetrate the gap. These states can significantly change the transport properties.

\clearpage

\bibliography{references}
	
\end{document}